\documentclass{aa}  

\usepackage{graphicx}
\usepackage{txfonts}
\usepackage{multirow}
\usepackage{longtable,lscape}
\usepackage{color}
\usepackage{multirow}
\usepackage{ccaption}

\definecolor{rojo}{rgb}{1,0,0}
\definecolor{blu}{rgb}{0,0,1}

\begin{document}

\title{A comprehensive study of NGC\,2345, a young open cluster with a low metallicity}
\titlerunning{The young open cluster NGC\,2345}

\author{
J. Alonso-Santiago\inst{1,2}
\and I. Negueruela\inst{3}
\and A. Marco\inst{2}
\and H. M. Tabernero\inst{2,4}
\and C. Gonz\'{a}lez-Fern\'{a}ndez\inst{5}
\and N. Castro\inst{6}
}

\institute{
INAF, Osservatorio Astrofisico di Catania, via S. Sofia 78, 95123, Catania, Italy\\
\email{javier.alonso@inaf.it} \\
\and Dpto de F\'{i}sica, Ingenier\'{i}a de Sistemas y Teor\'{i}a de la Se\~{n}al, Escuela Polit\'ecnica Superior, Universidad de Alicante, Carretera de San Vicente del Raspeig s/n, 03690, Spain \\
\and Dpto de F\'{i}sica Aplicada. Universidad de Alicante, Carretera de San Vicente del Raspeig s/n, 03690, Spain \\
\and Centro de Astrobiología (CSIC-INTA), Carretera Ajalvir km 4, E-28850 Torrejón de Ardoz, Madrid, Spain\\
\and Institute of Astronomy, University of Cambridge, Madingley Road, Cambridge CB3 OHA, UK\\
\and Leibniz-Institut für Astrophysik Potsdam, An der Sternwarte 16, 14482 Potsdam, Germany
}

\date{}

 
  \abstract
   {NGC\,2345 is a young open cluster hosting seven blue and red supergiants, low metallicity and a high fraction of Be stars which makes it a privileged
   laboratory to study stellar evolution.}
   {We aim to improve the determination of the cluster parameters and study the Be phenomenon. Our objective is also to characterise its seven evolved stars
   by deriving their atmospheric parameters and chemical abundances.}
   {We performed a complete analysis combining for the first time $ubvy$ photometry with spectroscopy as well as $Gaia$ Data Release 2. We obtained spectra with classification purposes for 
   76 stars and high-resolution spectroscopy for an in-depth analysis of the blue and red evolved stars.}
   {We identify a new red supergiant and 145 B-type likely members within a radius of 18.7$\pm$1.2\,arcmin, which implies an initial mass, $M_{\textrm{cl}}\approx$5\,200\,M$_{\sun}$.
   We find a distance of 2.5$\pm$0.2\,kpc for NGC\,2345, placing it at $R_{\textrm{GC}}$=10.2$\pm$0.2\,kpc. Isochrone fitting supports an age of 56$\pm$13\,Ma, implying masses around
   6.5\,M$_{\sun}$ for the supergiants. A high fraction of Be stars ($\approx$10$\%$) is found.
   From the spectral analysis we estimate for the cluster an average $v_{\textrm{rad}}$=$+58.6\pm0.5$\,km\,s$^{-1}$ and a low metallicity, [Fe/H]=$-$0.28$\pm$0.07. We also have
   determined chemical abundances for Li, O, Na, Mg, Si, Ca, Ti, Ni, Rb, Y, and Ba for the evolved stars. The chemical composition of the cluster is consistent with that 
   of the Galactic thin disc. One of the K supergiants, S50, is a Li-rich star, presenting an A(Li)$\approx$2.1. An overabundance of Ba is found, supporting the enhanced $s$-process.}
   {NGC\,2345 has a low metallicity for its Galactocentric distance, comparable to typical LMC stars. It is massive enough 
   to serve as a testbed for theoretical evolutionary models for massive intermediate-mass stars.}

   \keywords{open clusters and associations: individual: NGC~2345 -- Hertzsprung-Russell and C-M diagrams -- stars: abundances -- stars: fundamental parameters -- 
stars: late-type -- stars: emission-line, Be}

   \maketitle
%

\section{Introduction}\label{intro}
Stellar clusters are the natural laboratories to study stellar evolution, since all their members were formed from the same interstellar cloud. Because of this, cluster stars share distance,
age and initial composition, reason for which their observed evolutionary states will be strongly conditioned by their initial mass. 
Unlike globular clusters, where stellar evolution can be easily inferred since all the evolutionary branches are represented, open clusters instead offer us only short
snapshots because they are younger and less populated. In many cases young open clusters do not host evolved stars or just a few and, consequently, observations do not sufficiently
constrain theoretical models.

The transition between stars that explode as a supernova (SN) and those that do not, to date, still presents a great uncertainty. At solar metallicity, models set this 
limit around 10\,M$_{\sun}$ \citep{po08,doe15} whereas this value could decrease down to $\approx$7\,M$_{\sun}$ in binary systems \citep{pod04} and low-metallicity environments
\citep{pum09,ibe13}. If so, the number of SNe would be more than double what was supposed so far. In order to shed some light about this topic we have started an observational programme to study, 
in a consistent way, open clusters with ages comprised between 30--100\,Ma. 
With the aim of collecting a statistically significant sample that allows us to cover the AGB/RSG mass transition, we have selected for our study those clusters sufficiently massive to be representative,
hosting the largest number of evolved stars in this age range \citep{ignacio17,salamanca}. 
We expect to provide experimental results which can be used to improve theoretical models and be able to better constrain the physics of the most 
massive intermediate-mass stars. This is the third paper in the series after those devoted to NGC\,6067 \citep{6067} and NGC\,3105 \citep{3105}.

NGC\,2345 is a poorly studied cluster in the constellation of Canis Major [$\alpha$(2000)\,= 07h\:08m\:18s, $\delta$(2000)\,=\,$-13^{\circ}\:11\arcmin\:36\arcsec$; $\ell=226\fdg58$, $b=-2\fdg31$]. 
\citet{mof74} performed the first study of the cluster employing $UBV$ photoelectric photometry for 64 stars in the region defined by NGC\,2345 and slit spectra for the six brightest member stars. He found the cluster 
to host seven bright giants, two of which have spectral type A and the remaining five, type K. He found a variable reddening across the cluster ranging, in terms of $E(B-V)$, from 0.47 to 1.16. 
He also estimated a distance of 1.75 kpc and an age around 60 Ma, based on the earliest spectral type observed (around B4). In that study, 
by counting stars on a $V$-plate down to $V$=14 he estimated the presence of 42 members within the cluster, whose diameter was set at 10.5$\arcmin$. This value
was slightly increased up to 12.3$\arcmin$ (hosting 238 members) when the count was extended to fainter stars, down to a limiting magnitude around $V$=17.

Further studies on NGC\,2345 are scarce. \citet{be_2345} looking for Be stars in open clusters by employing slitless spectroscopy found that NGC\,2345 has the highest fraction of Be stars among all the clusters contained in their sample.
\citet{carr15} took the first and only $UBVRI$ CCD photometry to date, covering a field of view of 20$\arcmin$ x 20$\arcmin$ and reaching the 20th magnitude in the $V$ band. 
They determined an age for the cluster, assuming solar metallicity, in the range 63--70 Ma and a diameter of 7.5\arcmin. They placed the cluster at 3 kpc, doubling the 
\citet{mof74} value. 

From high-resolution spectra, \citet{reddy16} provided, for the first time, stellar atmospheric parameters as well as chemical abundances for three evolved members of NGC\,2345. 
Recently, \citet{hol19} performed a more detailed chemical analysis of the all five red giants found by \citet{mof74}. Both studies found a metallicity
for NGC\,2345 around [Fe/H]=$-$0.3, a very low value given its location within the thin disc of the Galaxy.

In this work we present the first Str\"{o}mgren photometry of NGC\,2345. In addition, we collect the largest collection to date of spectra for stars in the cluster field.
These spectra were taken at different resolutions with several instruments over more than six years. For the first time we provide spectra for the blue stars, including the bright A giants, not
studied since \citet{mof74} observed them a long time ago. We characterised these blue stars obtaining their atmospheric parameters and chemical abundances as well.
Additionally we also have observed the red giants at high resolution performing a detailed spectroscopic analysis.
Finally, we complement our own observations with $Gaia$ DR2 data in order to strengthen our analysis and be able to perform the most robust multi-approach study of this cluster so far.


\section{Observations and data}

\subsection{Photometry}

We obtained $ubvy$ Str\"{o}mgren photometry using the Wide Field Camera (WFC) on the 2.5-m Isaac Newton Telescope (INT) at the El Roque de los Muchachos Observatory, located in the Canarian island of La Palma (Spain). 
The WFC \citep{wfc} is mounted at the prime focus of the INT. It is an optical mosaic camera consisting of 4 thinned AR coated EEV 2k\,x\,4k CCDs. Each CCD has a pixel scale of 0.33$''$\,pixel\,$^{-1}$. 
The edge to edge limit of the mosaic is 34.2$'$, although in the inter-chip spacing around 1$'$ is neglected.  

The observations were acquired during the nights of 2011 December 23--25. For each band, we used different exposure times (see Table \ref{log}) in order to avoid saturated stars as well 
as cover a broad magnitude range. 

Reduction was done using a modified version of the pipeline developed at CASU for the IPHAS survey \citep[see][and references therein]{iphas}, as it uses the same instrumentation and a similar filterset. 
Briefly, the pipeline follows the standard optical reduction cascade, performing bias and dark subtraction followed by flat correction and gain homogenization over all four detectors.

Photometry was obtained using {\scshape imcore}, from the casutools suite\footnote{\url{http://casu.ast.cam.ac.uk/surveys-projects/software-release}}. Briefly, {\scshape imcore} performs object detection based on a robust 
estimation of the statistical properties of the background of the image, looking for connected groups of pixels above a user selectable threshold. For these detections, it obtains optimal aperture photometry
along with a curve-of-growth aperture correction to homogenize the measurements. Photometric calibration was done using all the repeated observations (both of calibration fields and target cluster) in a simplified
ubercal algorithm \citep{padman08}.

\begin{table}
\caption{Log of the photometric observations for NGC\,2345 taken at the INT.}
\begin{center}
\begin{tabular}{lccc}   
\hline\hline
\multirow{2}{*}{NGC\,2345} & \multicolumn{3}{c}{RA\,=\,07$^h$08$^m$18$^s$  DEC\,=\,$-13\degr11\arcmin36\arcsec$}\\
 & \multicolumn{3}{c}{(J2000) \,\,\,\,\,\,\,\,\,\,\,\,\,\,\,\,\,\,\,\,\,\,\,\,\,\,\, (J2000)}\\
\hline
\multirow{2}{*}{Filter} & \multicolumn{3}{c}{Exposure times (s)}\\
 & Short & Intermediate  & Long \\
\hline
$u$ & 100 & 600 & 1800 \\
$v$ & 10  & 300 & 1000 \\
$b$ & 2   & 60  & 600 \\
$y$ & 2   & 20  & 100 \\
\hline
\end{tabular}

\label{log}
\end{center}
\end{table}

The instrumental photometry (denoted with the subscript $i$) was calibrated employing selected standard stars observed the same nigths in the clusters NGC\,884 and NGC\,1039 \citep[the 
reference and the individual photometry for these stars are shown in][]{7419}. The equations used for computing the photometric transformation, according to \citet{ec_stromg}, are:

\begin{multline}
V=y_i+(-0.003\pm0.012)+(-0.04\pm0.05)\,\cdot\,(b-y)
\end{multline}

\begin{multline}
(b-y)=(-0.008\pm0.008)+(0.99\pm0.03)\,\cdot\,(b-y)_i
\end{multline}

\begin{multline}
m_1=(0.019\pm0.015)+(0.93\pm0.05)\,\cdot\,m_{1_i}+\\
+(-0.09\pm0.05)\,\cdot\,(b-y)
\end{multline}

\begin{multline}
c_1=(-0.070\pm0.014)+(1.049\pm0.012)\,\cdot\,c_{1_i}+\\
+(0.29\pm0.04)\,\cdot\,(b-y)
\end{multline}

In total, we have obtained photometry for 2153 stars, which are represented as blue circles in the first finding chart (Fig.~\ref{2345_grande}). 
In Table\,\ref{phot_2345} we list the identification number of each star ($ID$), the
equatorial coordinates (RA, DEC), the photometric values ($V$, $(b-y)$, $c_1$ and $m_1$) together with their uncertainties and the number of measurements ($N$). 
The magnitudes and colours are the averages, weighted with the variances, of all individual measurements. The errors are expressed in terms of the standard deviation around this mean. 

\subsection{Spectroscopy}

We obtained spectra in the field of NGC\,2345 for 76 stars in several runs, from 2010 to 2016,  with different spectrographs (see Table~\ref{spectra_2345}). Our sample consists of photometric cluster members according
to \citet{mof74}, 32, as well as early-type stars and supergiants in the surroundings of the cluster in order to study its extent.
Firstly, we selected those early-type stars previously catalogued as luminous stars \citep{LS} or Be stars \citep{SS77}. We also searched for red (super)giants according to their magnitudes and colours in 2MASS. 
Below we describe each run. 

In August 2010, during the nights of the 28th and the 29th, we took 126 spectra for 74 different stars by using the AAOmega spectrograph (AA$\Omega$) for classification purposes.
It is mounted on the 3.9-m Anglo-Australian Telescope (AAT) at the Australian Astronomical Observatory (AAO), on Siding Spring Mountain (Australia). 
This instrument is a dual-beam system, with two arms covering the full wavelength range, 3700--9500\,\AA{},
via a dichroic beam splitter with crossover at 5700\,\AA{}. AA$\Omega$ is an optical spectrograph that can operate in two modes: as a Multi-Object System (MOS) or as an Integral Field Unit (IFU).
In MOS mode, the setup used for us, AA$\Omega$ is fed by the Two DegreeField (2dF) fiber-positioning system, at the prime focus of the AAT. In this configuration 392 fibers are available, covering
a wide fov of 2$\degr$. 
The red arm was equipped with the 1700D grating (reaching a resolving power, $R\approx$10\,000 around the Ca triplet) whereas the blue arm used the grating 580V 
($R$=1\,300). The data reduction was performed using the automatic pipeline {\scshape 2dfdr} provided by the AAT as described in more detail in \cite{Ca15}. 
The detector is a 2k\,x\,4k E2V CCD.

In February 2011, from the 24th to the 27th, with IDS we obtained spectra for eight blue stars with the object of performing the spectral analysis to 
determine their stellar atmospheric parameters as well as their chemical abundances. IDS is mounted on the 2.5-m Isaac Newton Telescope (INT) at El Roque
de los Muchachos Observatory, in La Palma (Spain). 
It is a long-slit spectrograph placed at the Cassegrain focus and is equipped with a 235-mm focal length camera. The detector used is the default one, RED+2, a 4k\,x\,2k CCD red-sensitive and low-fringing array. 
We used the grating H1800V, at different central wavelengths, covering the spectral range around 3800--5380\,\AA{} with a $R\approx$5\,400.

On November 9, 2011 we reobserved in service mode with ISIS six of the stars previously observed with IDS with the aim of obtaining high-quality spectra to analyse. ISIS is mounted at the Cassegrain focus of the 4.2-m William Herschel Telescope (WHT) in La Palma. 
It is a high-efficiency, double-armed, medium-resolution spectrograph, capable of long-slit work up to $\approx4\arcmin$ length 
and $\approx22''$ slit width. Because of the use of dichroic filters simultaneous observing in both arms is possible. ISIS is equipped with two 4k\,x\,2k CCDs, the thinned and blue-sensitive EEV12 
(13.5 micron) on the blue arm and RED+ (15 micron) on the red one, a red-sensitive with almost no fringing camera. 
We used the grating R1200B centred around 4500\,\AA{}, which provides a $R\approx$3\,500.

We used WYFFOS for collecting spectra of 41 stars during three nights in September 2012, 15--17. 
WYFFOS works at the prime focus of the 4.2-m WHT. It is 
a multi-object fiber-fed spectrograph, containing 150 science fibers plus another 10 fiducial ones. Fibres are positioned by the Autofib2 system. WYFFOS covers a wide field of 40$\arcmin$.
We employed the grating R600R covering the spectral range between 6200--9200\,\AA{} at a resolution, $R\approx$1\,700, with the aim of identifying the sources presenting H$\alpha$ emission.
The camera (WHTWFC) is a mosaic consisiting of two thinned AR coated EEV 4k\,x\,4k CCDs. 

Finally, we resorted to FEROS in order to observe two early likely blue members and, in three different epochs, the cluster red (super)giants at high resolution which allow us to perform a detailed spectral analysis.
FEROS is mounted on the ESO/MPG 2.2-m telescope at La Silla Observatory (Chile). It is an \'{e}chelle spectrograph and covers the wavelength range from 3\,500\,\AA{} 
to 9\,200\,\AA{}, providing a resolving power of R=48\,000. The spectral range is covered in 39 orders, with small gaps between the orders appearing only at the longest wavelengths. FEROS is fed 
by two fibers that provide simultaneous spectra of the astrophysical target plus either sky or one of the two calibration lamps. Its detector is an EEV 2k\,x\,4k CCD.
In a first run during May 2011, 11--13, under ESO programme 087.D-603(A), we took spectra for eight stars. 
Then, a second run with FEROS was carried out in May 2015, 29--30 for six stars under ESO programme 095.A-9020(A). Finally, we downloaded from the ESO archive the spectra of five stars
taken with FEROS under ESO programme 096.A-9024(A) on March 11, 2016.

\begin{figure*}  
  \centering         
  \includegraphics[width=15cm]{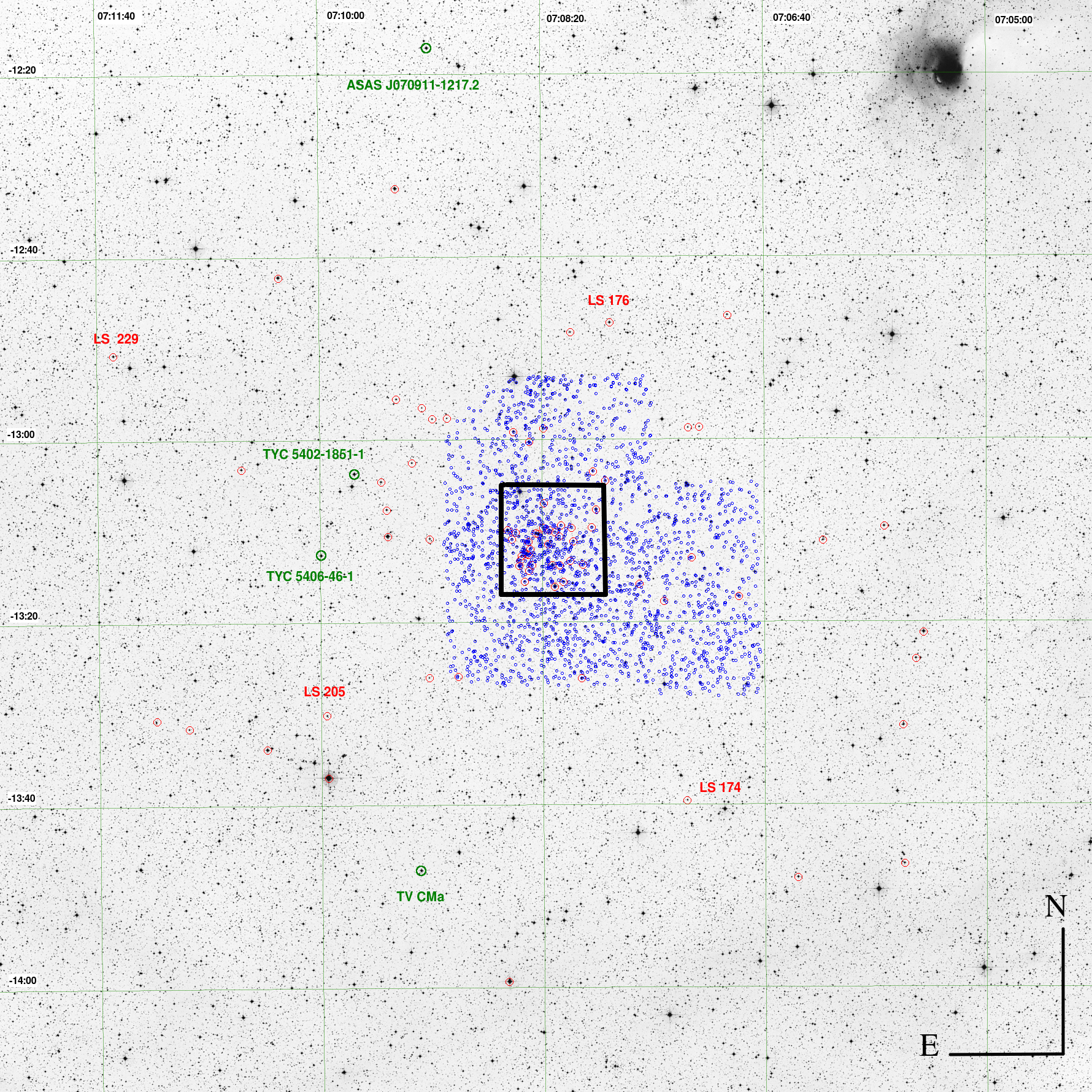}   
  \caption{Finding chart for stars with spectra (red circles) in the field and surroundings of NGC\,2345. The chart is a 120$'$x120$'$ POSS2 Red image. The stars covered by our photometry are identified as blue circles. 
  Some stars whose relationship with the cluster is discussed in the text have also been represented (green circles).The central part of the cluster, inside the black square, is shown in more detail in Fig.~\ref{2345}. North is up and East is left.}

  \label{2345_grande}  
\end{figure*}

\begin{figure*}  
  \centering         
  \includegraphics[width=15cm]{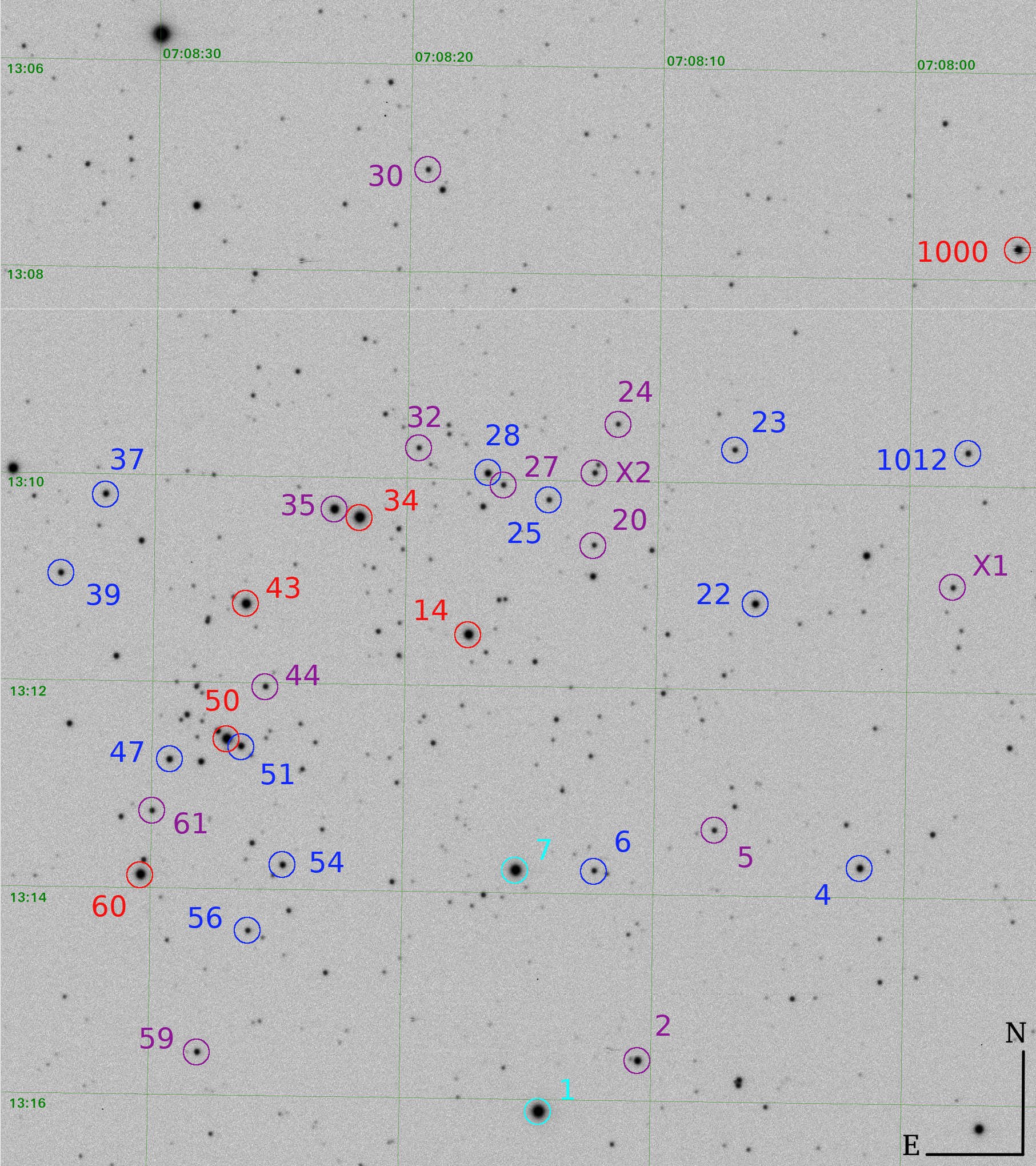}   
  \caption{Finding chart of the core of NGC\,2345, the black square in Fig\,\ref{2345_grande}. The image is one of our $V$-band frames and covers a region of sky around 10$'$x11$'$. Stars for which we have spectra are inside circles of different 
  colours according to their spectral types, as explained in the text, to facilitate its identification. Stars X1 and X2, identified as Be stars by \citet{mathew11}, without a WEBDA numbering, are also included. North is up and East is left.} 
  \label{2345}  
\end{figure*}

Figures \ref{2345_grande} and \ref{2345} show the stars for which we have spectra and photometry. In the first finding chart the region covered by our 
photometry is marked, as well as all the stars with spectroscopy, spread over a 
wide field centred on the cluster. The second chart, instead, is limited to the central part of NGC\,2345.

\subsection{Archival data}
We completed our observations with data from all-sky surveys, such as 2MASS and $Gaia$, inside a radius of $30\arcmin$ around the nominal cluster centre.
We took the $JHK_{\textrm{S}}$ magnitudes from the 2MASS catalogue \citep{2MASS} only for stars with good-quality photometry (i.e. without any ``$U$'' photometric flag in the catalogue).
In addition, we also employed the $Gaia$ DR2 data \citep{GaiaDR2} for those stars with sufficiently good astrometric data (i.e. a parallax error smaller than 0.5 mas).


\section{Results}\label{results_2345}

Throughout this paper we followed the WEBDA numbering (denoted as $Star$), which is the one established by \citet{mof74}. For those stars not 
observed by him we employed an arbitrary and sequential numbering beginning with 1000. When possible, we used the LS designation corresponding to stars listed 
in the catalogue of Luminous Stars in the Southern Milky Way compiled by \citet{LS}. The designation in our photometry (ID) can be found in Tables~\ref{phot_2345} and~\ref{phot_spt}.
Table~\ref{spectra_2345} lists both designations for stars with spectra.

\subsection{Spectral classification}\label{sp_2345}

We performed the spectral classification of the observed objects based on classical criteria detailed below. The results obtained are described depending on the
temperature range, i.e. hot (or blue) and cool (red) stars. Table~\ref{spectra_2345} shows for all the stars forming our sample their 
numbering, assigned spectral types and spectrographs with which they were observed. For three stars, because of the poor quality of their 
spectra, it was impossible to assign a spectral type. For all the rest, we estimate for our classification a typical uncertainty of around one spectral subtype. 

\subsubsection{Blue stars}

\begin{figure*} 
  \centering         
  \includegraphics[width=16cm]{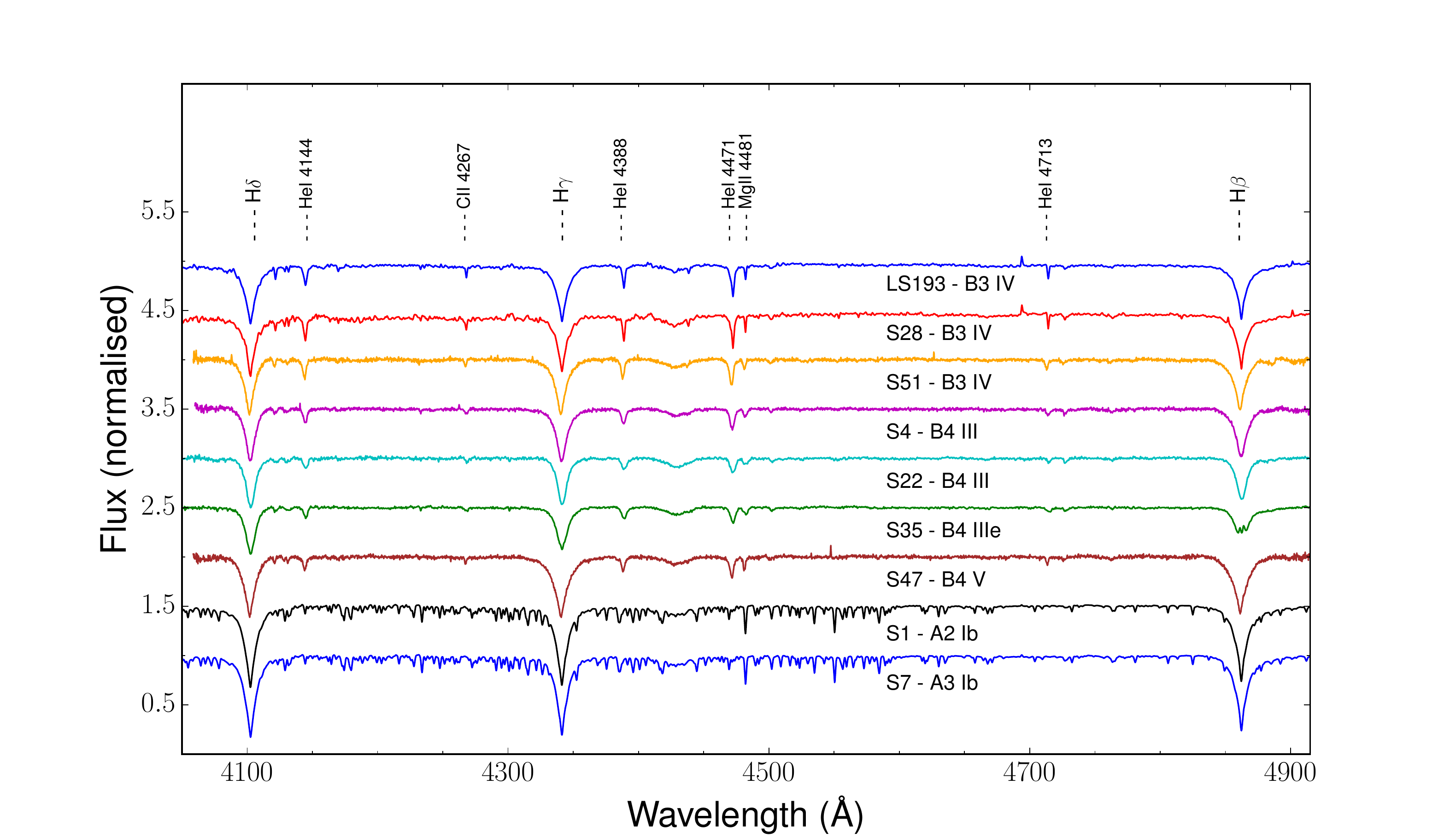}   
  \caption{Spectra of the brigthest blue stars in the field of the cluster, taken with ISIS (Stars 4, 47 and 51), IDS (22 and 35) and FEROS (1, 7, 28 and LS193). 
  The most important lines in this spectral region are marked. Notice the increasing width of the Balmer lines as we move from giants
  to main sequence among the mid-B stars.}
  \label{esp_az}  
\end{figure*}

We took spectra of the blue stars in the field of NGC\,2345 to study the upper main sequence (MS) and find which is the spectral type corresponding to the main sequence turnoff (MSTO) point
in the photometric diagram. 
We have four high- and 14 intermediate-resolution spectra for nine different stars, finding two A supergiants and seven other stars with early-mid B spectral types (see Fig.~\ref{esp_az}). 
We followed classical criteria of classification in the optical wavelength range (4\,000\,--\,5\,000\,\AA), following \citet{Ja87} and \citet{gray}.

With respect to the A supergiants, around spectral type A2--A3 the \ion{Ca}{ii}~K-line is a notable feature while the profile of the Balmer lines is the primary luminosity 
criterion: the narrower their wings are, the higher the luminosity. The \ion{Fe}{ii} line at 4233\,\AA{}, the blends of \ion{Fe}{ii} and \ion{Ti}{ii} at 4172\,--\,8\,\AA{} and the 
\ion{Si}{ii}\,$\lambda$4128\,--\,30 doublet are enhanced in the supergiants. 

Regarding the B-type stars, as we move toward later spectral types the ratio \ion{Mg}{ii}\,$\lambda$4481/\ion{He}{i}\,$\lambda$4471 increases, as \ion{He}{i} lines weaken. 
Our sample consists of stars with spectral types between B3 and B4. The line ratios used in this range to assign spectral types are \ion{Si}{ii}\,$\lambda$4128\,--\,30/\ion{Si}{iii}\,$\lambda$4553, 
\ion{Si}{ii}\,$\lambda$4128\,--\,30/\ion{He}{i}\,$\lambda$4121, \ion{N}{ii}\,$\lambda$3995/\ion{He}{i}\,$\lambda$4009, and \ion{He}{i}\,$\lambda$4121/\ion{He}{i}\,$\lambda$4144 \citep{Wa90}.
Among likely members (see Table~\ref{spectra_2345}) the earliest spectral type found on the main sequence is B3\,V.

Finally, we observed at low resolution, just for classification purposes, one O-, 40 B- and 12 A-type stars. Most of these objects are field stars located in the 
surroundings of the cluster, in the region not covered by our photometry (Fig.\,\ref{2345_grande}).

\subsubsection{Be stars}

We identified 17 Be stars in a wide field centred on NGC\,2345. Most of them (10) are located in the central part of the cluster (i.e. d\,$<\,5\,\arcmin$ from the nominal centre). We observed almost the same stars 
at a similar resolution, $R\approx$\,1\,500, in two different epochs separated roughly by two years. In 2010 we took spectra around H$\beta$ with AA$\Omega$ whereas in 2012 with WYFFOS we focused on H$\alpha$.  
Eleven of these 17 Be stars showed emission in both epochs whereas star LS\,171 only in H$\alpha$ in 2012. The remaining five objects were observed only one time. Table~\ref{tab_be} lists the Be stars
found in this work together with their spectral types and distances from the cluster centre.
In addition, for S35 we also have spectra in two epochs during 2011 around H$\beta$ (taken with IDS and ISIS) in which emission was observed.

\begin{figure}  
  \centering         
  \includegraphics[width=\columnwidth]{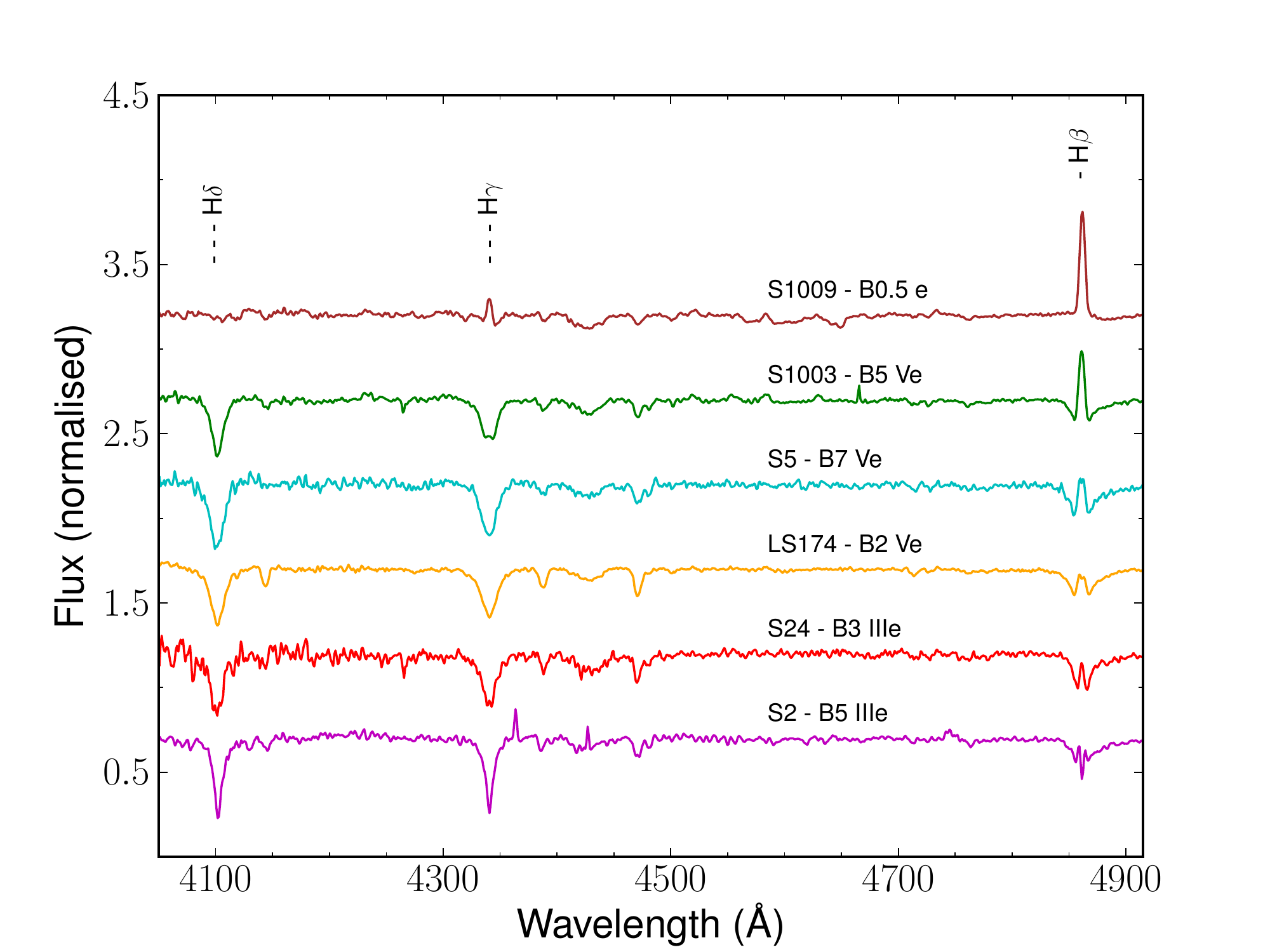}   
  \caption{AA$\Omega$ spectra of some Be stars in the field of NGC\,2345. As a reference, Balmer lines are marked. Stars 1009 (the only non member in this figure) and 1003 stand out especially for their strong emission. In the former
  star the emission phenomenon is also observable in H$\gamma$. } 
  \label{be}  
\end{figure}

For some of these Be stars, spectra in the blue region are represented in Fig.~\ref{be}. Notice the strong emission of stars 1003 and 1009, especially the last one, which also exhibits H$\gamma$ in emission.
These two stars, together with S2 \citep[also mentioned in][]{Me82b} and S1125, are known Be stars included in the catalogue of Galactic emission stars by \citet{SS77}. S1125 is the only one among them without 
any emission feature in our spectra (although it only has been observed around H$\beta$, it could still show emission in H$\alpha$). In \citet{mof74} is not mentioned the presence of any Be star in the cluster.
\citet{be_2345} carried out a survey to identify likely Be stars in 207 young open clusters using slitless spectroscopy. 
Later, \citet{mathew11} observed and confirmed at low resolution, around H$\alpha$ and H$\beta$, those objects found in the previous work as Be-star candidates. In NGC\,2345 they identified
12 Be stars, out of which ten have a WEBDA numbering (included in Table~\ref{tab_be}). We have in common with them nine objects (only S27 was not observed by us), whereas we found emission in S30 and S63 (not quoted 
in their work). Two other Be stars found by \citet{mathew11} without a WEBDA numbering are not included in our spectroscopic sample (both are identified in the finding 
chart, Fig.~\ref{2345}).

\begin{table}
\caption{Spectral types and distances to the cluster centre for the Be stars found in the field of NGC\,2345 in this work in H$\alpha$ (2012) and H$\beta$ (2010). The Be stars, identified by \citet{mathew11}, Mat11, are added
for comparison as well as those listed with their numbering (SS) according to \citet{SS77}. The ``$+$'' symbol indicates emission whereas a ``$-$'' represents the lack of it.
\label{tab_be}}
\begin{center}
\begin{tabular}{lcccccc}   
\hline\hline
\multirow{2}{*}{Star} & \multirow{2}{*}{Sp T} & \multirow{2}{*}{d\,($\arcmin$)} & \multicolumn{2}{c}{This work} & \multirow{2}{*}{Mat11} & \multirow{2}{*}{SS} \\
&  &  & H$\alpha$& H$\beta$&  & \\
\hline
\multicolumn{7}{c}{Members}\\    
\hline
2       & B5\,IIIe & 4.4  & $+$ & $+$ & $+$ & 87 \\
5       & B7\,Ve   & 3.1  &     & $+$ & $+$ &    \\
20      & B5\,Ve   & 1.7  &     & $+$ & $+$ &    \\
24      & B3\,IIIe & 2.6  & $+$ & $+$ & $+$ &    \\
27      & ---      & 1.6  &     &     & $+$ &    \\
30      & B7\,IIIe & 4.6  & $+$ & $+$ &     &    \\
32      & B5\,Ve   & 1.9  &     & $+$ & $+$ &    \\
35      & B4\,IIIe & 1.7  & $+$ & $+$ & $+$ &    \\
44      & B5\,Ve   & 1.9  & $+$ & $+$ & $+$ &    \\
59      & B5\,Ve   & 4.7  &     & $+$ & $+$ &    \\
61      & B7\,Ve   & 3.3  & $+$ & $+$ & $+$ &    \\
63      & B5\,Ve   & 12.8 & $+$ & $+$ &     &    \\
1003    & B5\,Ve   & 9.5  & $+$ & $+$ &     & 86 \\
\hline
\multicolumn{7}{c}{Halo members}\\    
\hline
LS\,174 & B2\,Ve   & 31.8 & $+$ & $+$ &     &    \\
LS\,176 & B3\,IIIe & 25.3 & $+$ & $+$ &     &    \\
\hline
\multicolumn{7}{c}{Non members}\\    
\hline
1009    & B0.5e    & 37.0 & $+$ & $+$ &     & 82 \\
1051    & B5\,III shell    & 47.9 &     & $+$ &     &    \\
1125    & B5\,III  & 42.3 &     & $-$ &     & 81 \\
LS\,171 & B1.5\,Ve & 32.0 & $+$ & $-$ &     &    \\
\hline
\end{tabular}
\end{center}
\end{table}

\subsubsection{Red stars}

We took high-resolution spectra for seven evolved stars as well as low-resolution spectra for other four FG stars in the cluster field. For classification, we focused on the 
spectral region around the \ion{Ca}{ii} triplet in the near-infrared wavelengths (8480--8750\,\AA{}). The triplet weakens toward later spectral types and lower 
luminosity classes \citep{Ja87}, but many other classification criteria are available in this range. Many features of \ion{Fe}{i} (i.e. lines at 8514, 8621 and 8688\,\AA{}) 
and \ion{Ti}{i} (8518\,\AA{}) become stronger with later spectral types \citep{carquillat}. In addition, we find very useful as classification criteria the ratios 
\ion{Ti}{i}\,$\lambda$8518\,/\,\ion{Fe}{i}\,$\lambda$8514 and \ion{Ti}{i}\,$\lambda$8734\,/\,\ion{Mg}{i}\,$\lambda$8736, which become larger with increasing spectral type.

\begin{figure*}  
  \centering         
  \includegraphics[width=16cm]{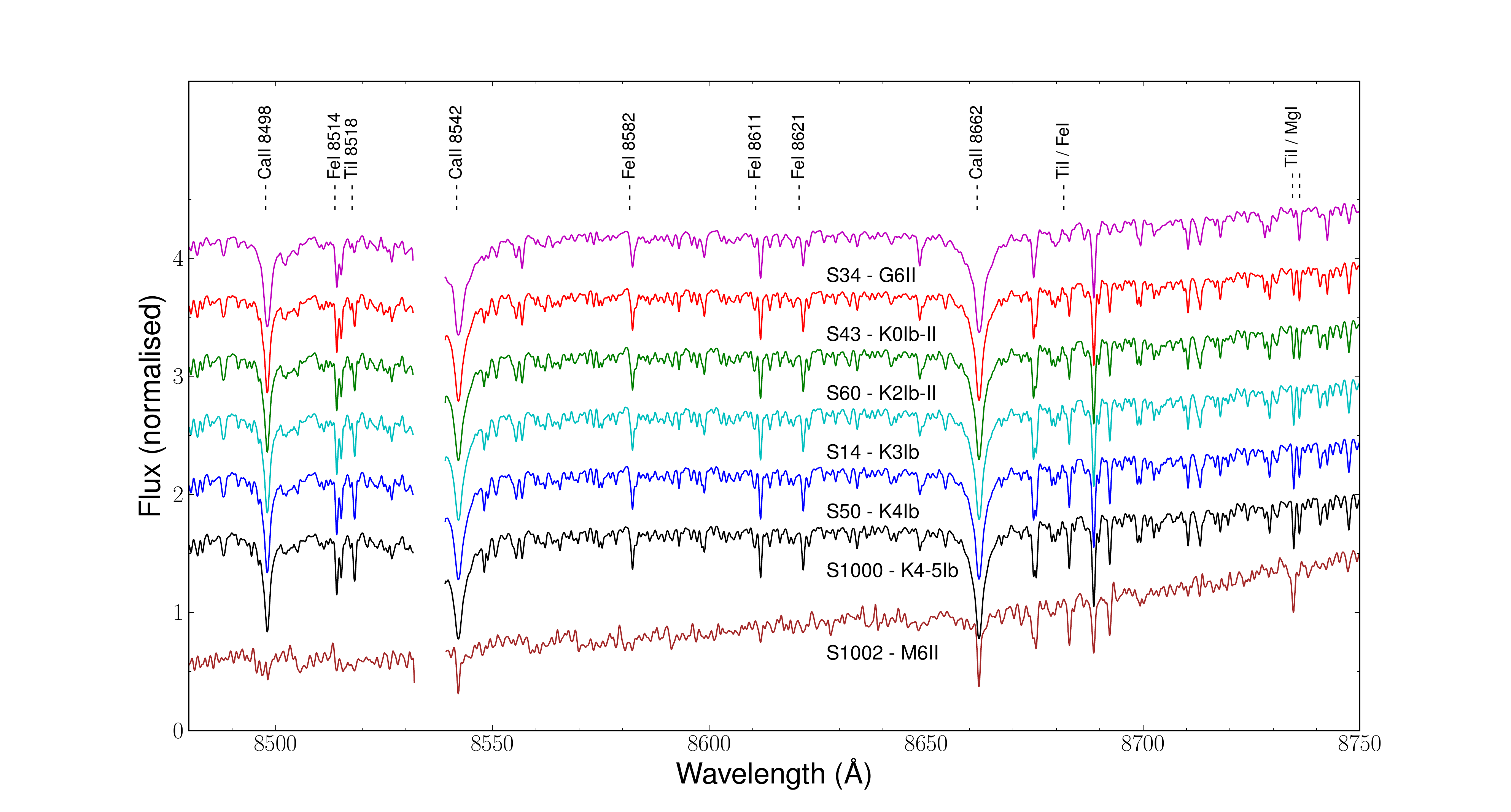}   
  \caption{FEROS spectra of the red supergiants observed around the \ion{Ca}{ii} triplet. Note the gap between orders 37 and 38 between 8532--8540\,\AA{}. 
  S1000 is the new K supergiant found in this work.} 
  \label{esp_rojos}  
\end{figure*}

Figure~\ref{esp_rojos} displays spectra of the evolved stars observed at high resolution, which are classified as G (1), K (5) and M (1) bright giants/low-brightness supergiants.
Star 34 is a binary star composed of a red giant (G6\,II) which dominates the composite spectrum and a blue companion (B:) which is only slightly noticeable 
by the weak exhibition of its Balmer lines. Besides the five RSGs found in this cluster by previous authors \citep{mof74,Me08,hol19}, we identify 
two possible candidates to be included among the cluster members: S1000 (a K supergiant) and S1002 (a M-type object), this latter detected in the cluster halo and thus not covered by 
our photometry. Both stars (IRAS sources 07055-1302 and 07068-1321, respectively) were targeted based on their positions on the 2MASS CMD and their RVs, obtained with AA$\Omega$ in our
first run, compatible with those of the other evolved members.

\subsection{Radial and rotational velocities}\label{rv_2345}
We determined heliocentric radial velocity (RV) for the spectroscopically observed stars through Fourier cross-correlation by employing different softwares in function of the type of stars. 
For blue ones we used the {\scshape fxcor} task included in the {\scshape iraf} package.\footnote{{\scshape iraf} 
is distributed by the National Optical Astronomy Observatories, which are operated by the Association of Universities for Research in Astronomy, Inc., under cooperative agreement
with the National Science Foundation.}.
We correlated the spectrum of each star against a template and the correlation peak was 
fitted with a Gaussian. The templates were selected among known stars covering the B spectral type from a previous work \citep{6067}. We focused our attention on the range between
4\,000 and 5\,000\,\AA{}. In order to improve the correlation the diffuse interstellar band around 4428\,\AA{} was masked as well as the region around H$\beta$ when emission is strong. 
Conversely, in the case of cool stars we employed {\scshape iSpec} \citep{ispec}, computing the cross-correlation against a list of atomic lines masks, carefully selected for the {\em Gaia}
benchmark stars library pipeline, from asteroids observed with the NARVAL spectrograph.

For objects with moderate- (IDS) and high-resolution (FEROS) spectra we determine accurate RVs displayed in Tables~\ref{Par_cal_2345} and \ref{Par_2345}. For the remaining stars their RVs derived 
from the AA$\Omega$ spectra are listed in Table~\ref{spectra_2345}. Most of them correspond to blue stars that have only been observed with this instrument. Since the resolution used is low ($R$=1\,300)
our errors in the cross-correlation are of the order of 30\,km\,s$^{-1}$ (around 10$\%$ of instrumental broadening). For this reason the RV listed for these stars should be considered only as an 
indicative value of their real RV. 

From red giants (i.e. S14, S43, S50 and S60), excluding S34 because of its binarity, we estimate the mean RV for the cluster by averaging the individual values, obtaining a $v_{\textrm{rad}}$=$+58.5\pm0.5\,\textrm{km}\,\textrm{s}^{-1}$,
where the error is the dispersion between the values for the different stars. In the same way \cite{Me08} and \cite{hol19} computed similar RVs: 59.2$\pm$0.7\,km\,s$^{-1}$ and 58.5$\pm$0.4\,km\,s$^{-1}$, respectively.
Instead, \citet{reddy16} used stars 34, 43 and 60 deriving a $v_{\textrm{rad}}$\,=\,59.7\,$\pm$\,3.4\,km\,s$^{-1}$. The large dispersion is produced by star 34, a binary object 
whose RV is somewhat larger than the rest. If we do the same we obtain a value in fair agreement with that, 60.1\,$\pm$\,3.5\,km\,s$^{-1}$.

Based on the average RV for the cluster we selected as likely members those stars with a RV compatible with it. All the stars observed with FEROS, according to their RVs, are considered as likely members,
including stars whose cluster membership had never been studied, namely S1000, S1002 and LS\,193. Only stars 28 and 34 show RV somewhat different from the other stars. The latter one can be explained because of
its binarity. The same explanation might be valid for S28, since its behaviour throughout the analysis carried out in this paper is fully compatible with membership. 
Among the members identified by \citet{mof74} all of them share a RV (derived from AA$\Omega$ spectra) compatible with the cluster average. Only 
S36, whose $v_{\textrm{rad}}\approx-6$\,km\,s$^{-1}$  is clearly an outlier and, thus, we rejected its membership to the cluster.
Taking into account the final selection of members (see Sect.~\ref{membership}) we updated the cluster RV by including all single members (blue and red, 12 in total) observed with FEROS and IDS.
The resulting value is $v_{\textrm{rad}}$\,=58.3\,$\pm$\,1.7\,\,km\,s$^{-1}$, where the error cannot be considered as a real estimate of the RV dispersion of the cluster. 
It is rather the scatter originated by the larger uncertainty when computing RV from blue stars (observed at lower resolution) instead of red (super)giants.

For 11 stars observed with FEROS we determined projected rotational velocity ($v \sin\,i$) by using the {\scshape iacob-broad} code \citep{iacob_broad}, based on the Fourier transform method, which separates rotation
from other broadening mechanisms such as the macroturbulent velocity ($\zeta$). For hot stars, as a diagnostic, we used the line of \ion{Mg}{ii} at 4481\,\AA{} while for cool stars we employed, at least, eight
lines of \ion{Fe}{i} and \ion{Ni}{i}. In addition, for five other stars, all of them blue objects, we only have IDS spectra. Their resolution is not as high as that necessary to employ the usual method for separating
different broadenings. In this case we assumed that the whole broadening has a rotational origin. Starting from a first estimation of the $v\sin\,i$ (around 50\,km\,s$^{-1}$, a value close to the resolution) it was
computed a first model capable of reproducing the spectrum. In an iterative way, once we set the stellar parameters, we looked for a second estimate of rotation by changing it until finding a new model which 
reproduced best the profiles. This process was repeated a couple of times until it did not change the rotation. These velocities are listed in Table~\ref{Par_cal_2345} for blue stars and in Table~\ref{Par_2345} 
for the cool objects. The errors reflect the scatter between measurements, in terms of rms. 

\subsection{Cluster membership}\label{membership}
The disentanglement and identification of cluster members from field stars is not a straightforward task. The very precise astrometry provided by $Gaia$ DR2 
is the ideal tool to carry it out. In a first step we looked for the likely cluster members by resorting to the $Gaia$ DR2 astrometric data, i.e. parallax ($\varpi$) and proper motions ($\mu_{\alpha*}$, $\mu_{\delta}$).
From the $Gaia$ values for the members already known according to the literature \citep{mof74}, we placed the cluster in the space of the astrometric parameters. By employing 31 stars in the 
cluster core for which we have spectra (see Table~\ref{tab_mp}), we found the cluster at ($\varpi$, $\mu_{\alpha*}$, $\mu_{\delta}$)$_{cl}$\,=\,(0.35, $-$1.36, 1.33)\,$\pm$\,(0.03, 0.10, 0.09) after
discarding two field stars (S47 and S36, the latter in addition with a radial velocity totally incompatible with that of the cluster). The error ($\sigma$) is expressed in terms of the standard
deviation of the individual values. Our determination is very similar to that computed in an automatically way by \citet{cantat18} also using $Gaia$ data, (0.348, $-$1.332, 1.340). 

\begin{figure*}  
  \centering         
  \includegraphics[width=16cm]{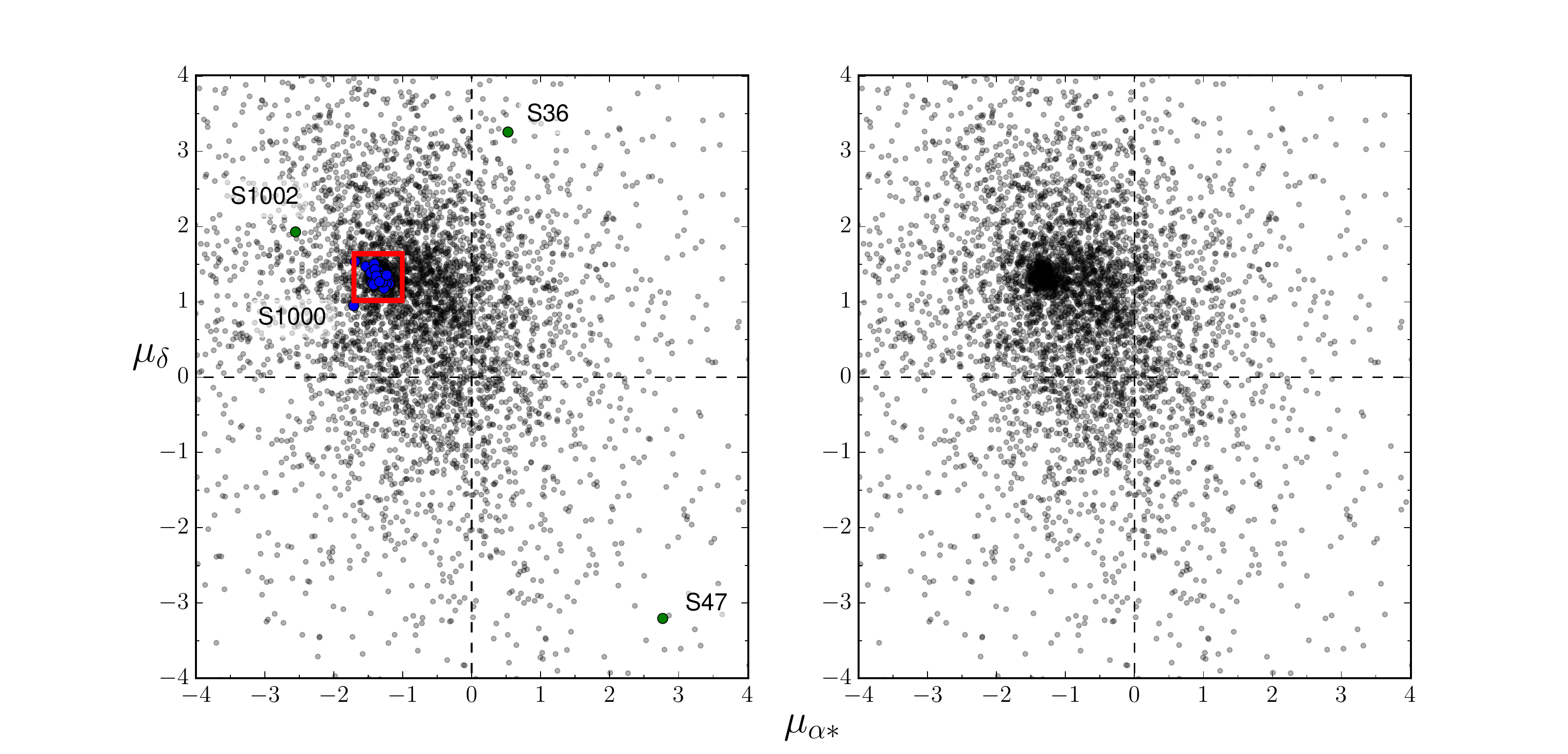}   
  \caption{Proper-motion diagram for NGC\,2345. \textbf{Left}: Grey dots are the $Gaia$ sources within 10$\arcmin$ around the nominal cluster centre whereas 
  coloured circles represent stars observed spectroscopically in this work: in blue, likely members previously cited in the bibliography and in green, field 
  stars. The red square draws the boundary of the cluster according to that described in the text. \textbf{Right}: For clarification purposes only $Gaia$ 
  sources are represented: the cluster clearly stands out from the background.} 
  \label{graf_mp}  
\end{figure*}

Once we obtained the ``cluster centre'' in the astrometric space, in a second step we defined the cluster as those stars inside a circle centred at ($\varpi$, $\mu_{\alpha*}$, $\mu_{\delta}$)$_{cl}$
within a 3-$\sigma$ radius. Additionally, in order to take into account the $Gaia$ uncertainties we added to this radius the average astrometric errors found for the known members (0.03, 0.05, 0.04).
In this way we identified 834 likely cluster members within the range 0.23$\leq$\,$\varpi$\,$\leq$0.47, $-$1.71$\leq$\,$\mu_{\alpha*}$\,$\leq-$1.01 and 1.02$\leq$\,$\mu_{\delta}$\,$\leq$1.64. 
In Fig.~\ref{graf_mp} the proper-motion diagram for the cluster is represented. On this diagram, the field stars S36 and S47 lie clearly outside the cluster boundary just defined.
In addition, the location of the two possible new (super)giants is shown. Both of them have RV fully compatible with that of the cluster (Sect.~\ref{rv_2345}) as well as
occupying a position on the 2MASS CMD similar to the rest of red members. Nevertheless, according to the astrometric data we are forced to discard the membership of S1002 to the cluster 
(0.4178$\pm$0.1514, $-$2.556$\pm$0.228, 1.936$\pm$0.196).
Conversely, S1000 is on the edge to be considered a member based on its astrometry (0.3932$\pm$0.0539, $-$1.712$\pm$0.078, 0.950$\pm$0.066). However, taking into account its 
astrometric uncertainties (greater than the average values) the star can be accepted as a cluster member. Additionally, we claim for its membership since from the spectroscopic analysis 
we do not find any significant differences with respect to the other red members already confirmed.

Finally, we have also evaluated the membership of the stars observed with AA$\Omega$ in the surroundings of the cluster. According to their RVs and, specially, 
their $Gaia$ data, we found that eight stars (namely S1003, S1010, S1012, LS\,174, LS\,176, LS\,193, LS\,205 and LS\,229) are compatible with the average values for the cluster. 
Morever, seven of these stars could be likely members when also taking into account that the their spectral types are in agreement with those observed in the cluster. 
The only exception is S1010, whose spectral type, BN2\,Ib, is not compatible with its membership. These seven stars are located in a wide range between 5$'$ and 52$'$ from the nominal
cluster centre. This fact could be an evidence of the existence of a large halo or dispersed association around the cluster.
	
\subsubsection{Selection of B-type members}\label{sec_miem}
In young clusters like NGC\,2345 B-type stars populate the upper MS, being the brightest cluster stars (obviously not taking into account the evolved members). 
These stars are key in our work since they serve us to estimate the reddening and the mass of the cluster. For this reason the next task 
consists of identifying the B-type stars among all the members previously found, for which the Str\"{o}mgren photometry is particularly suitable. We crossed 
our photometry with the likely members astrometrically selected, finding 365 objects in common \citep[361 when crossing it with the likely members
selected by][]{cantat18}. 

\begin{figure} 
  \centering         
  \includegraphics[width=\columnwidth]{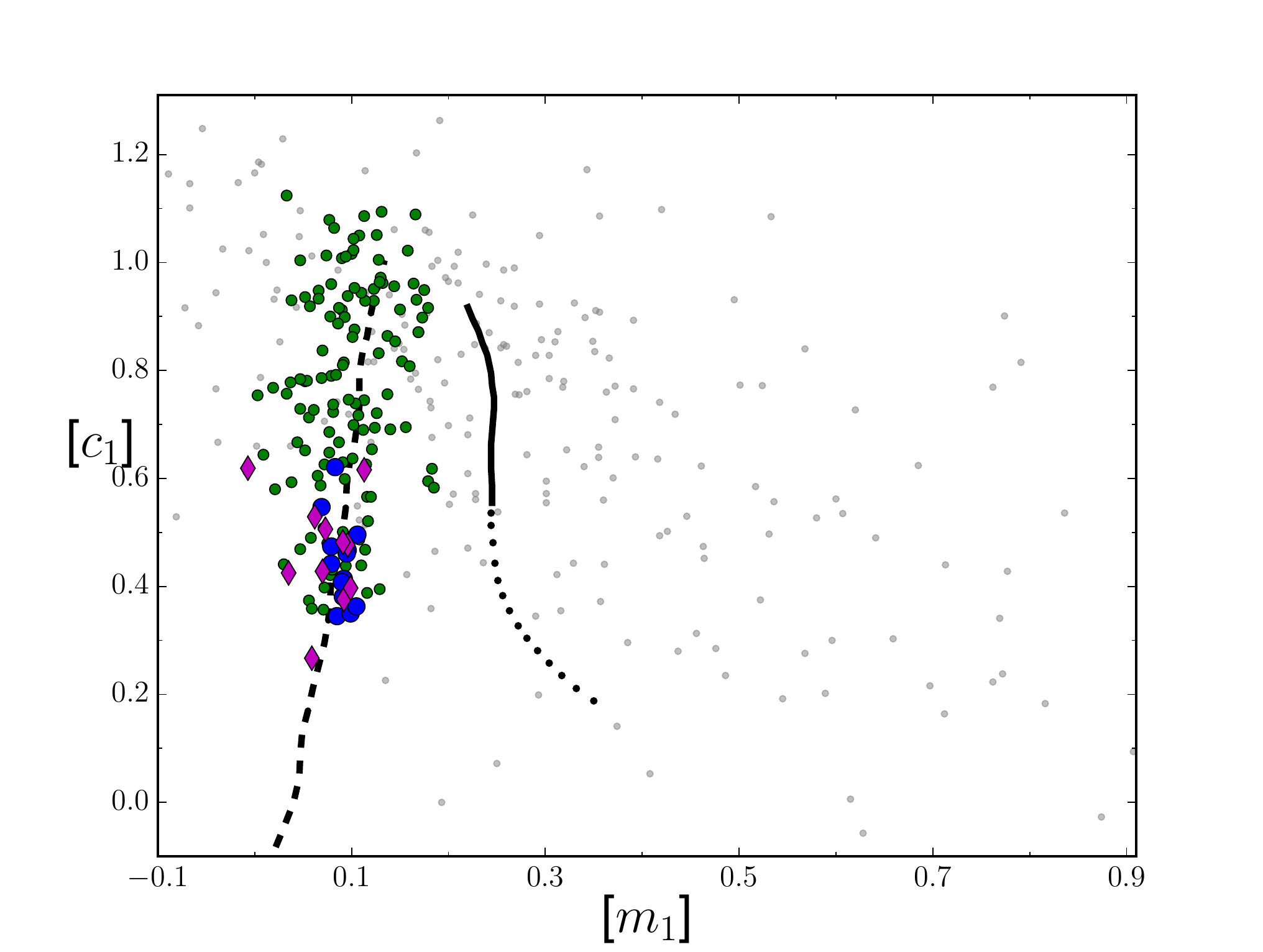}   
  \caption{$[c_1]/[m_1]$ diagram for likely members of NGC\,2345. The black lines represent the standard relation from \citet{perry87} for B stars (dashed line),
  A stars (solid line) and F stars (dotted line). B and Be stars observed spectroscopically are shown as blue circles and magenta diamonds, respectively. Green
  circles indicate those stars selected as B-type among all the remaining members (grey dots). }
  \label{c1_m1}
\end{figure}

\begin{figure} 
  \centering         
  \includegraphics[width=\columnwidth]{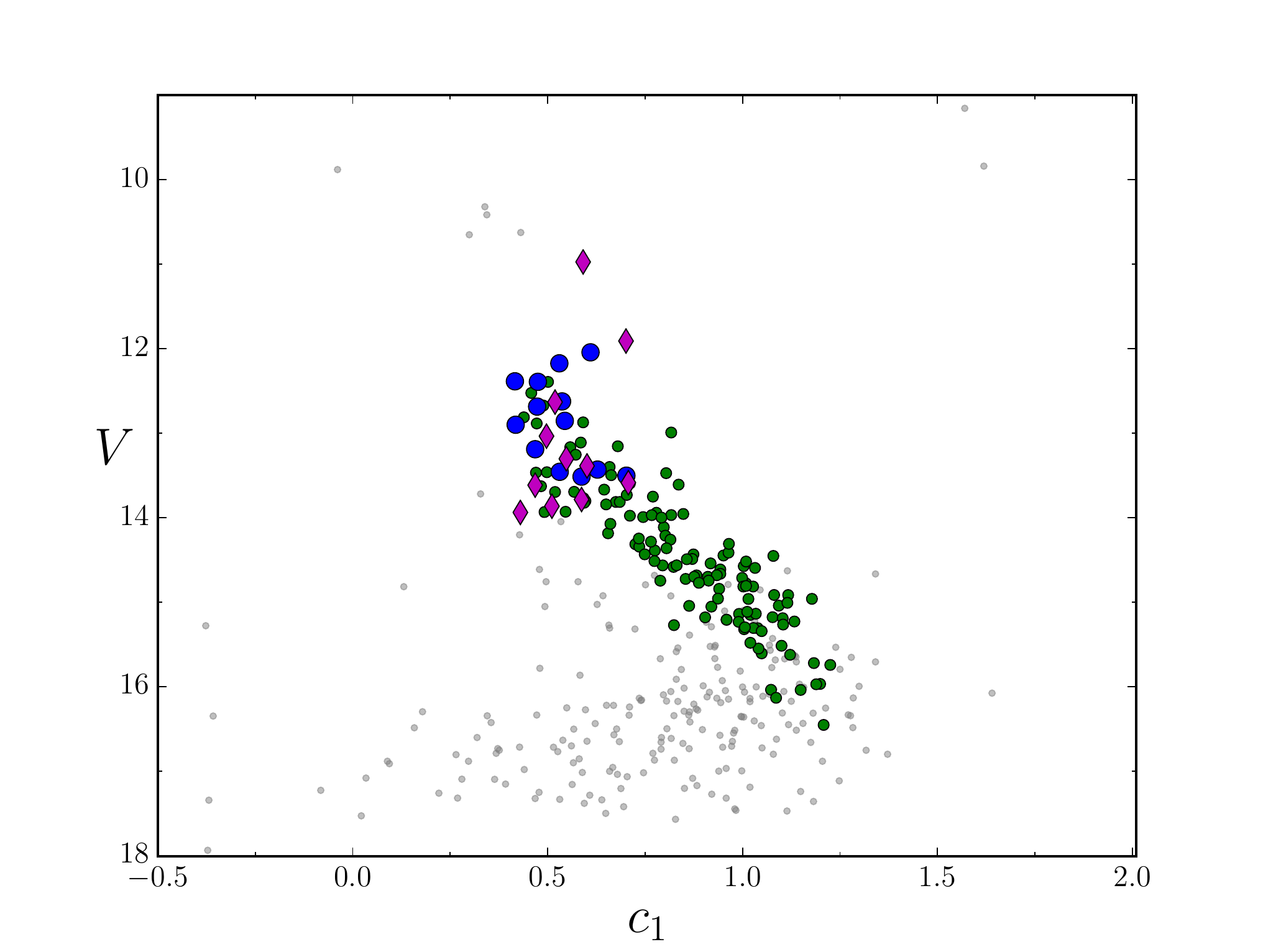}   
  \caption{$V/c_1$ diagram for likely members of NGC\,2345. Colours and symbols are the same as those in Fig.~\ref{c1_m1}.}
  \label{v_c1}
\end{figure}

Then, we plotted the [$c_1$]/[$m_1$] diagram\footnote{[$c_1$] and [$m_1$] are extinction-free indices computed as [$c_1$]=$c_1-0.20\cdot(b-y)$ and 
[$m_1$]=$m_1+0.32\cdot(b-y)$, respectively.} 
with the aim of searching B-type stars (Fig.~\ref{c1_m1}). We confirmed the reliability of our selection on a $V$/$c_1$ diagram, where all these stars are 
located following the same sequence, as expected (Fig.~\ref{v_c1}). In order to check our analysis, we overimposed in both diagrams those members observed spectroscopically, 
confirming that their location is in good agreement with their spectral types.
In this way we identified 122 B-type objects in addition to those, already known, observed with spectra: 13 B-type and 11 Be stars. 
In total, we found 145 likely B members in NGC\,2345.

\begin{figure} 
  \centering         
  \includegraphics[width=\columnwidth]{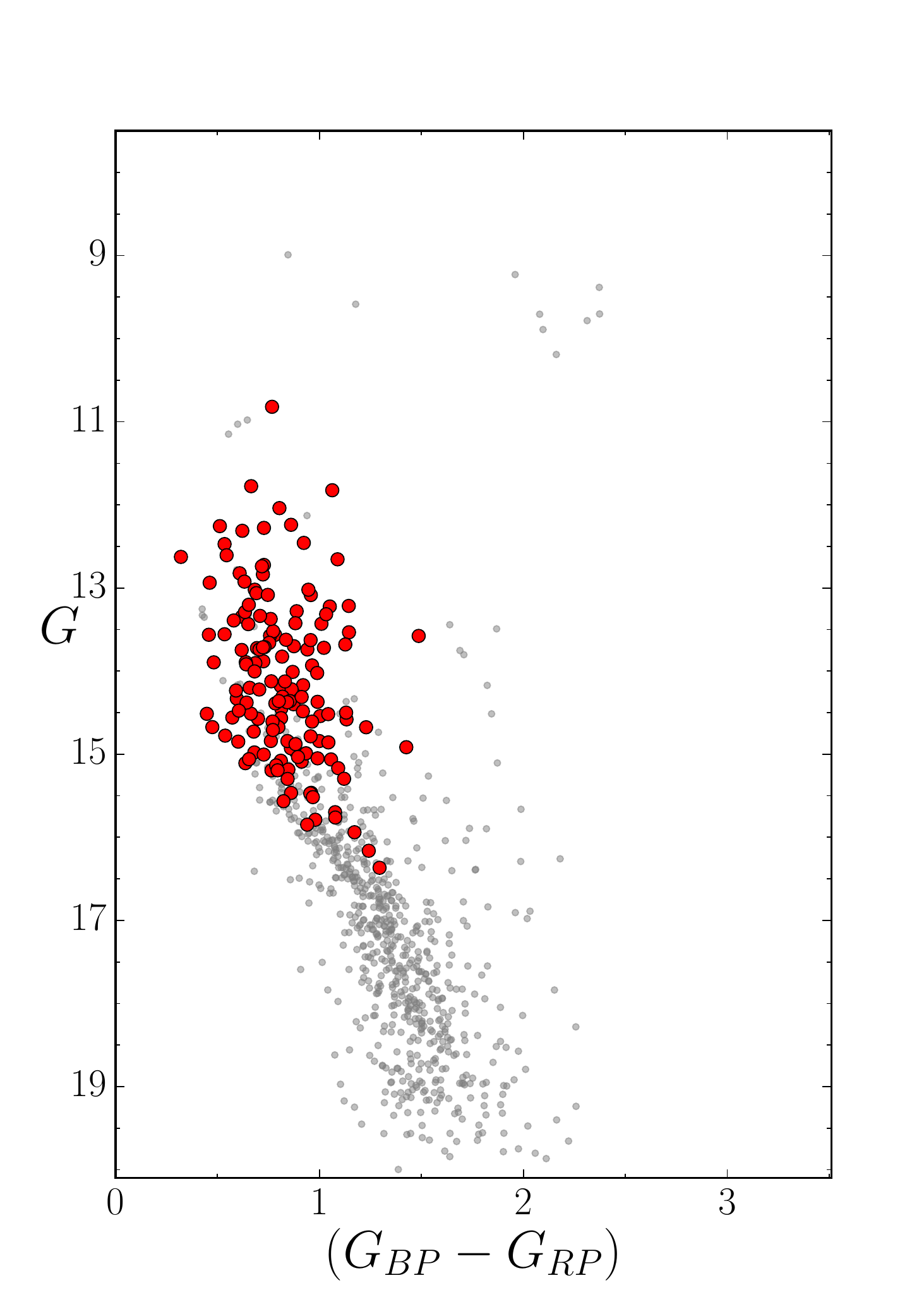}   
  \caption{$G/(G_{BP}-G_{RP})$ diagram for the stars in the field of NGC\,2345 whose astrometry is compatible with the cluster average values. Likely B members, 
  selected from our photometry, are shown as red circles.}
  \label{gaia_cmd}
\end{figure}

Finally, in order to check the impact of the field contamination in our member selection we plotted the Fig.\,\ref{gaia_cmd}. We displayed on a $Gaia$ CMD the 834 likely cluster 
members according to their astrometric properties and, among them, we highlighted those selected as B-type based on our photometric analysis. We do not appreciate an important 
effect that spoils our blue members selection. As expected, they are at the top of the cluster MS (clearly visible in the diagram), more than three magnitudes above the limiting value 
($\approx\,G=20$). Only a small dispersion is seen in their distribution, surely due to the differential reddening present across the cluster field. We note that those
bright members not included in our selection are B stars not covered by our photometry, including some LS stars such as LS\,176 (which is the brightest object among the three stars with $\approx\,G=11$)
and LS\,195 (with a magnitude slightly fainther than $G$=12). 

\subsection{Cluster reddening}
After evaluating the cluster membership, we estimated individual reddenings for likely B-type members assuming standard extinction laws, i.e. $E(c_1)=0.2\,E(b-y)$
and $E(B-V)=0.74\,E(b-y)$. The colour excesses in $(b-y)$ are computed following the procedure described by \citet{crawford70} in an iterative way. We used the observed $c_1$ to predict the first approximation to $(b-y)_0$ 
with the expression $(b-y)_0=-0.116+0.097c_1$. Then we calculated $E(b-y)$ and used the standard relation to correct $c_1$ for reddening, $c_0=c_1-E(c_1)$. The intrinsic colour $(b-y)_0$ was then calculated by replacing $c_1$
with $c_0$ in the above equation for $(b-y)_0$. After three iterations we finally obtain reliable individual $E(b-y)$. From the likely members we estimated the average cluster reddening, $E(b-y)=$\,0.49$\pm0.10$ corresponding to 
$E(B-V)=$\,0.66$\pm0.13$. In both cases, errors show the standard deviation of the individual values. 

The high dispersion suggests the presence of noticeable differential reddening across the observed field, as described by previous studies \citep{mof74,carr15}.
We found that the reddening varies from 0.41 to 1.13, reaching the highest values to the northwest, in the area approximately bounded by the stars 22, 23, 25 and 1012 (see Fig.\ref{2345}). 

\subsection{Determination of distance and age}
\begin{figure}
  \centering         
  \includegraphics[width=\columnwidth]{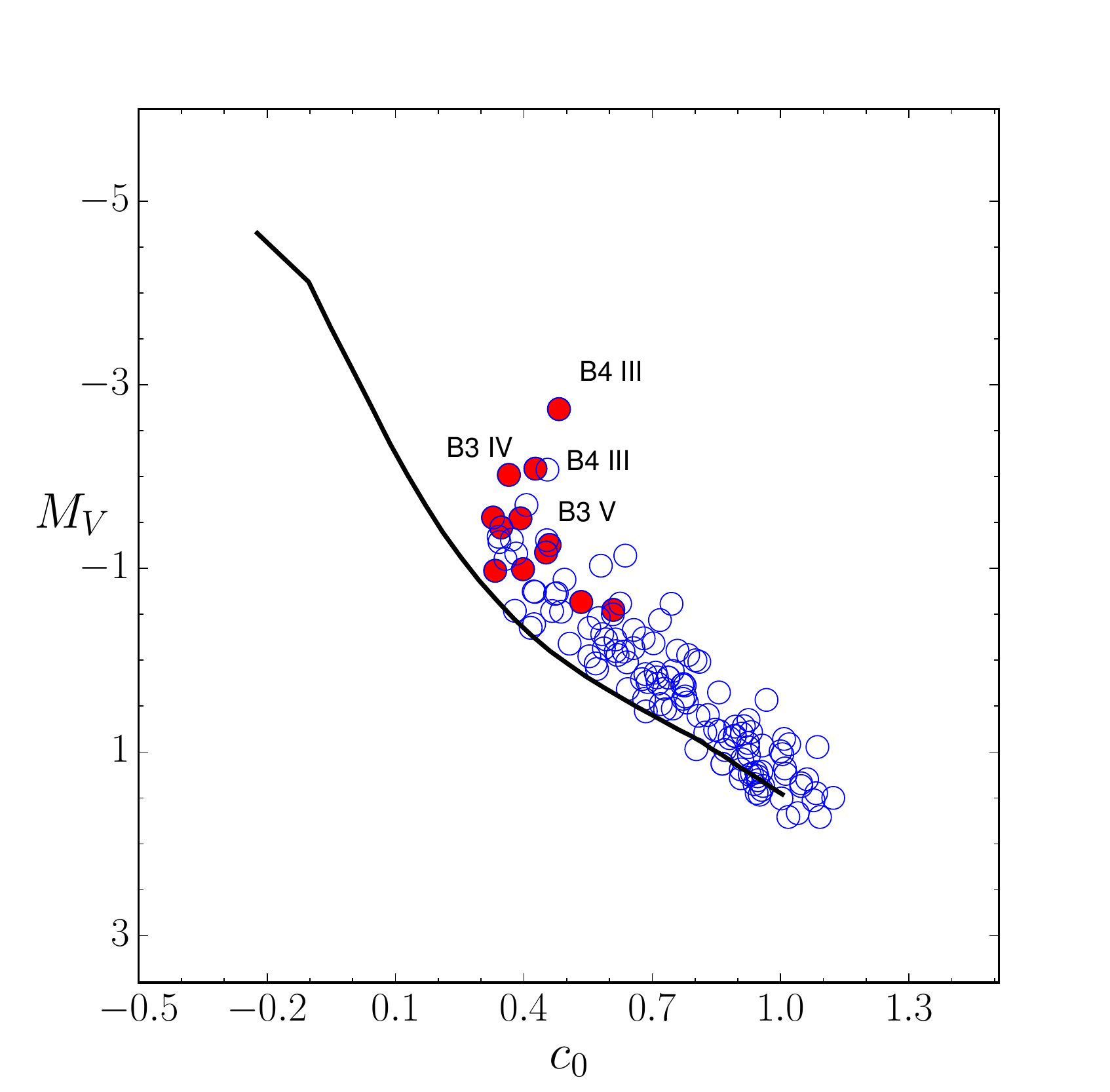}   
  \caption{$M_V/c_0$ diagram for blue likely members of NGC\,2345. Stars observed spectroscopically are represented in red colour. Spectral types for the brightest
  stars are marked. Blue open circles are the remaining early members selected photometrically. The black line shows the calibration of \citet{perry87} for B stars.}
  \label{zams}
\end{figure}

We continue the analysis of the cluster parameters by calculating the distance at which it is located. By the one hand, from the individual reddening for blue likely members, we carefully carried out a visual fit 
of these dereddened values to those calibrated by \citet{perry87} in the $M_V/c_0$ diagram (Fig.~\ref{zams}). In this way we obtain the distance modulus, $\mu=V_0-M_V=12.0\pm0.2$, which corresponds to a distance of $d=2.5\pm0.2$\,kpc. The error reflects
the uncertainty when visually fitting the ZAMS as a lower envelope. The combination of photometry and spectroscopy offers us the possibility of checking the validity of our
result. In this way, stars for which we have spectra show absolute magnitudes according to their spectral types.

By the other hand, we check this estimate obtained from the photometric analysis with the value computed from astrometry. 
The cluster mean parallax derived from the 145 B likely members ($\varpi=0.349\pm0.039$\,mas) is fully compatible with that found by \citet{cantat18}, 0.348, in their astrometric
analysis, where they quoted a distance of 2.65\,kpc. Thus, our value is consistent with the $Gaia$ parallax.
This distance places the cluster, with respect to the Galactic centre and taking as solar reference $R_{\sun}=8.3\,$kpc, at $R_{\textrm{GC}}=10.2\pm0.2\,$kpc.

\begin{figure}
  \centering         
  \includegraphics[width=\columnwidth]{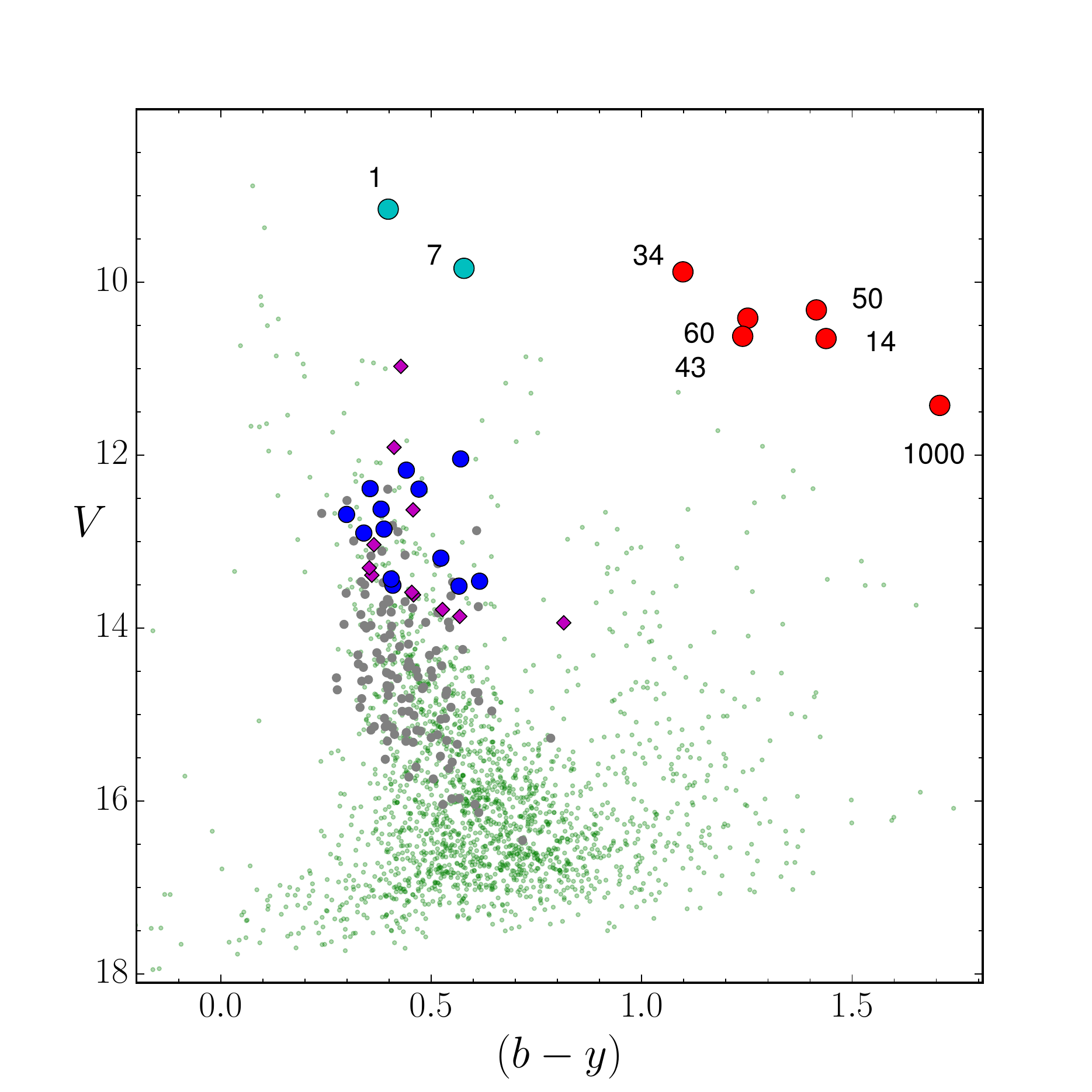}   
  \caption{$V/(b-y)$ diagram for all stars in the field of NGC\,2345. Photometric data appear as green dots. Grey dots represent likely B members. 
  Stars observed spectroscopically are represented as red circles (RSGs), cyan circles (A supergiants), blue circles (blue stars) and magenta diamonds (Be stars).} 
  \label{CMD}
\end{figure}

\begin{figure*}
  \centering         
  \includegraphics[width=17cm]{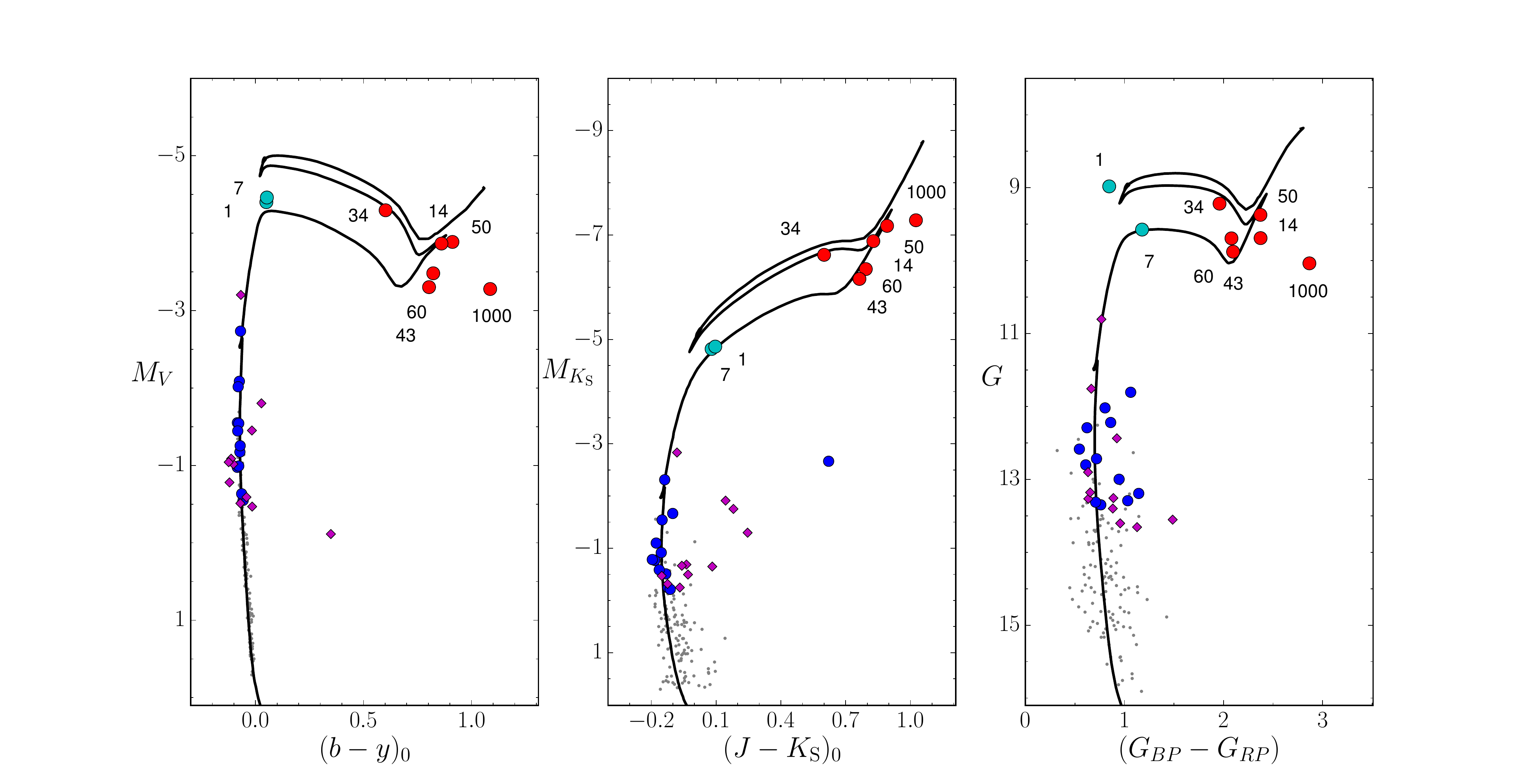}   
  \caption{Colour-magnitude diagrams for likely members of NGC\,2345 in three different photometric systems: \textbf{Left:} $M_V/(b-y)_0$ from the Str\"{o}mgren photometry taken in this work;  
  \textbf{Center:} $M_{K_{\textrm{S}}}/(J-K_{\textrm{S}})_0$ (2MASS photometry) and \textbf{Right:} $G/(G_{BP}-G_{RP})$ ($Gaia$ DR2 data). Photometric data for likely members appear as grey dots.
  Stars observed spectroscopically are represented as red circles (RSGs), cyan circles (A supergiants), blue circles (blue stars) and magenta diamonds (Be stars). The black line shows the best-fitting
  PARSEC isochrone (log\,$\tau$=7.75).}
  \label{isocronas}
\end{figure*}

Once we estimated the cluster distance we employed the isochrone fitting method to determine its age. 
Firstly, with the intention of showing the location of the evolved stars, we plot the $V/(b-y)$ colour-magnitude diagram (CMD) for all stars in the field (Fig.~\ref{CMD}). Before drawing the dereddened CMD on which we will trace isochrones, 
it is necessary to deredden the remaining members, namely Be stars and blue and red (super)giants. Taking into account the differential reddening we do not use 
the mean value found for the cluster. Instead, we compute an individual reddening for each source by averaging those of the closest B members. The thus made dereddened $M_V/(b-y)_0$ CMD is displayed 
in Fig.~\ref{isocronas} (left panel). The decrease in the stellar colour dispersion attests to the dereddening sucess.

In addition, in Fig.~\ref{isocronas} two other CMDs are also shown: $M_{K_{\textrm{S}}}/(J-K_{\textrm{S}})_0$ (central panel) and $G/(G_{BP}-G_{RP})$ (right panel). 
In the latter case, since the dereddening of the $Gaia$ photometry is not a trivial task we have preferred to redden the isochrone \citep[with the corrected $Gaia$ passbands from][]{maiz18}.
On each CMD we plot several isochrones corresponding to different ages, shifted to the distance modulus previously obtained. The best-fitting isochrone was carefully chosen, by eye, 
paying special attention to the position of the evolved stars. We adopted a PARSEC isochrone \citep{parsec} with the metallicity derived for the cluster in this work, [Fe/H]=$-0.28$ (See Sect.~\ref{feh}). 
We converted this value into $Z$ by taking into account the aproximation [M/H]=log($Z/Z_{\odot}$), with $Z_{\odot}$=0.0152 for PARSEC tracks. 

As can be seen in Fig.~\ref{isocronas} the 2MASS CMD yields the best fit since IR bands are less affected by the dust extinction. In this CMD all the evolved members lie quite well on the isochrone.
Only S1000 does not macht it as well as the other red giants. Morever, it is an IRAS source (numbering as 07055-1302) which suggest an IR excess. Taking into account this fact, 
the actual position of these stars could be compatible with that of an E-AGB star. In the two other CMDs, however, the fitting of the evolved stars is not so good probably due to the 
difficulty of correcting propertly the variable reddening across the cluster field.
The best-fitting isochrone is the same in all the three CMDs, log\,$\tau=7.75\pm0.10$, from which we estimate an age for the cluster of 56$\pm$13\,Ma.
The error shows the range of isochrones that give a good fit. At this age the mass of the red (super)giants corresponds to $\approx$\,6.5\,$M_{\sun}$.

\subsection{Size and mass of the cluster}

The centre of a cluster is the region where the stellar concentration is higher. By detecting this overdensity it is therefore possible to estimate the cluster centre.
For this task we resort to the likely astrometric members previously selected. We compute the stellar density profile by counting stars along each equatorial coordinate (RA, DEC).
By fitting one Gaussian to each profile, and assuming a bin size defined according to the Freedman--Diaconis rule, we find the maximum stellar density at $\alpha$=107.085$\pm$0.007\,deg and $\delta$=$-$13.197$\pm$0.008\,deg.
Significant differences with the nominal centre are not appreciated (this work$-$nominal), $\Delta\alpha$=0.0$\pm$0.4\,arcmin and $\Delta\delta$=0.2$\pm$0.5\,arcmin.

\begin{figure} 
  \centering         
  \includegraphics[width=\columnwidth]{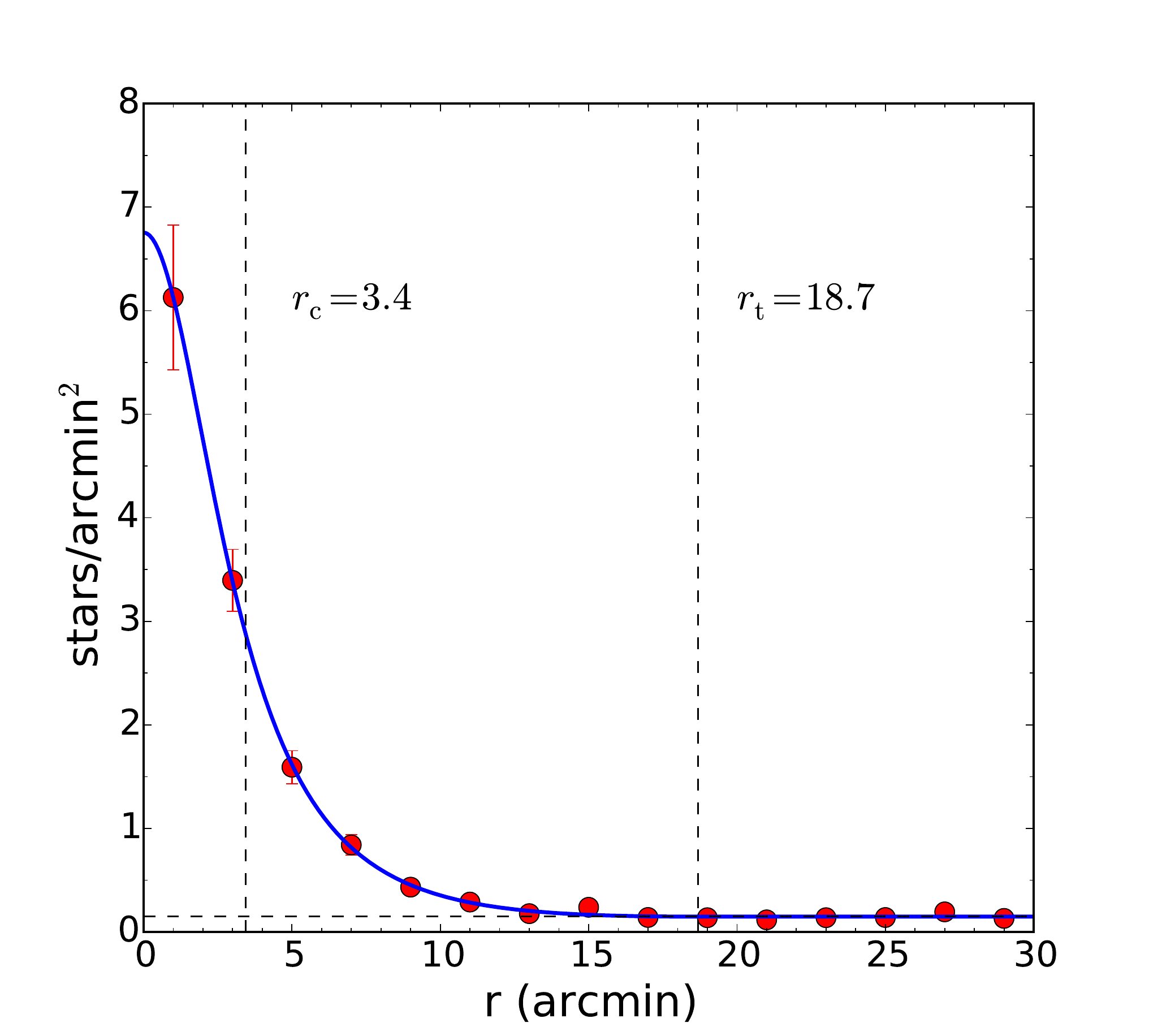}   
  \caption{Projected density distribution of the likely $Gaia$ DR2 members in the the cluster field. The red circles are the observed values together with their Poisson errors. The blue line
  shows the fitted King profile from which we obtained both the core ($r_{\textrm{c}}$) and the tidal ($r_{\textrm{t}}$) radii. Their positions are indicated with vertical dashed lines. The background density (0.15 stars\,arcmin$^{-2}$)
  is also represented as a horizontal dashed line.}
  \label{king}
\end{figure}

We determine the cluster size from the evaluation of the stellar projected distribution around its nominal centre. The stellar density profile, $\rho(r)$, was then estimated from star 
counts in concentric annuli around the cluster centre up to a reasonable distance of 30 arcmin (see Fig.~\ref{king}). Finally, we fit this density profile to a three-parameters King-model \citep{king}:

\begin{equation}
 $ $\rho$(r) = $\rho$$_0$ $\left\{\displaystyle\frac{1}{\sqrt{1+(r/r_c)^2}} - \displaystyle\frac{1}{\sqrt{1+(r_t/r_c)^2}} \right\}^2$ $
\end{equation} 

where $\rho_0$ is a constant, $r_{\textrm{c}}$ is the cluster core radius and $r_{\textrm{t}}$ is the tidal radius. The core radius is defined as the radial distance at which the value of density
becomes half of the central density and the tidal one as the distance at which the cluster disappears in its environment. 
The values obtained from the fitting are $r_{\textrm{c}}$=3.44$\pm$0.08\,arcmin and $r_{\textrm{t}}$=18.7$\pm$1.2\,arcmin, which correspond at the distance of the cluster to physical sizes of
$r_{\textrm{c}}$=2.63$\pm$0.24\,pc and $r_{\textrm{t}}$=14.3$\pm$1.9\,pc, respectively.
This value is consistent with the position of the most remote RV members (i.e. S64, S1003 or S1020), located at $\approx10\arcmin$ from the cluster centre.
In addition, all the B likely members are inside the cluster extent. 
The value calculated in this work is larger than that estimated by \citet{Pi08}, $r_{\textrm{t}}$=9.7$\pm$2.6\,pc.

We estimate the mass of NGC\,2345 following the same procedure as in the analysis of NGC\,3105 \citep[][where the reader can find a detailed explanation]{3105}.
Once we know the cluster size we select all the B members inside this radius (in this case including all the early members previously selected). In order to estimate the cluster mass we employed 
the multiple-part power law initial mass function (IMF) defined by \citet{Kroupa}. In a first step we set the free parameter of the IMF by counting the stars within a certain mass range were we have 
completeness. We chose stars located at the top of the MS, from spectral type B3\,V to B9\,V, in total 145 objects (previously selected in Sect.~\ref{sec_miem}). 
According to the isochrone, this range of spectral types covers stars between 2.6 and $6.9\,$M$_{\sun}$. 
In a second step, from this free parameter and integrating the IMF we obtain the cluster mass. After correcting from binarity we obtained a present cluster mass of 
$3\,700\,$M$_{\sun}$, which is equivalent to an initial mass of $\approx$\,5\,200\,M$_{\sun}$. This implies that NGC\,2345 is a moderately massive cluster, similar to other young open clusters (such as NGC\,6067, NGC\,3105 and Be\,51).

\subsection{Stellar atmospheric parameters}\label{parametros}

We computed stellar parameters for 16 stars following the procedures described in the analysis of NGC\,3105 \citep{3105}. 
We divided the sample in two different groups: the early stars and the late stars. We used, among the available spectra for each star, the one with the highest resolution. 
There are 11 FEROS spectra (four blue and seven cool) and five other blue objects taken with IDS. Results are displayed in Tables~\ref{Par_cal_2345} (for blue
stars) and \ref{Par_2345} (for the cool ones).

\subsubsection{Blue stars}

We computed stellar parameters for nine blue stars from their IDS and FEROS spectra.
We used a grid of {\scshape fastwind} synthetic spectra \citep{Si11,Ca12}. The stellar atmosphere code {\scshape fastwind} \citep{Sa97,Pu05} enables non-LTE 
calculations and assumes a spherical geometry as well as an explicit treatment of the stellar wind effects. 
The technique employed was described by \citet{Ca12}, also in \citet{Le07}. The stellar atmospheric parameters were derived through an automatic $\chi^2$-based 
algorithm searching for the set of parameters that best reproduce the main strong lines observed in the range $\approx$\,4000\,--\,5000\,\AA{}. 
The results (i.e. effective temperature, surface gravity and macroturbulent velocity) are shown in Table~\ref{Par_cal_2345}. 
We note that the temperatures obtained are in good agreement with the values expected from the calibration by \citet{Hump84}.

\begin{table*}
\caption{Log of FEROS and IDS spectra for blue stars and stellar atmospheric parameters derived from them. \label{Par_cal_2345}}
\begin{center}
\begin{tabular}{lccccccccc}   
\hline\hline
Star & Sp T & $t_{\textrm{exp}}$ (s) & $S/N$ & $v_{\textrm{rad}}$ (km\,s$^{-1}$) & $v \sin\,i$ (km\,s$^{-1}$) &  $\zeta$ (km\,s$^{-1}$) & $T_{\textrm{eff}}$ (K) & $\log\,g$ \\
\hline
\multicolumn{8}{c}{FEROS}\\
\hline
1      & A2\,Ib    & 1\,350 & 160 & 55.57 $\pm$ 0.25 & 25.3 $\pm$ 0.2 & 22.4 $\pm$ 0.7 &  8\,250 $\pm$ 1\,000 & 1.9 $\pm$ 0.4 \\
7      & A3\,Ib    & 1\,500 & 150 & 60.12 $\pm$ 0.21 & 24.1 $\pm$ 1.0 & 24.7 $\pm$ 3.3 &  8\,250 $\pm$ 1\,000 & 2.0 $\pm$ 0.3 \\
28     & B3\,IV    & 1\,000 & 70  & 49.06 $\pm$ 0.57 & 29.7 $\pm$ 2.7 & 32.0 $\pm$ 2.0 & 16\,000 $\pm$ 1\,400 & 3.7 $\pm$ 0.2 \\ 
LS\,193& B3\,IV & 1\,000 & 70  & 56.96 $\pm$ 0.61 & 43.2 $\pm$ 1.2 & 15.5 $\pm$ 3.2 & 16\,000 $\pm$ 1\,300 & 3.7 $\pm$ 0.1 \\
\hline
\multicolumn{8}{c}{IDS}\\
\hline
4  & B4\,III & 3\,600 & 80 & 62.16 $\pm$ 0.98 & 170 & --- & 15\,000 $\pm$ 1\,000 & 3.3 $\pm$ 0.1 \\
22 & B4\,III & 3\,600 & 80 & 56.89 $\pm$ 1.44 & 210 & --- & 14\,000 $\pm$ 1\,000 & 3.0 $\pm$ 0.1 \\
35 & B4\,IIIe& 3\,200 & 110& 57.52 $\pm$ 0.24 & 210 & --- & 15\,000 $\pm$ 1\,000 & 3.0 $\pm$ 0.1 \\ 
47 & B4\,V   & 1\,800 & 80 & 56.30 $\pm$ 1.69 & 90  & --- & 15\,000 $\pm$ 1\,000 & 3.7 $\pm$ 0.1 \\
51 & B3\,IV  & 3\,600 & 80 & 57.67 $\pm$ 2.00 & 110 & --- & 16\,000 $\pm$ 1\,000 & 3.5 $\pm$ 0.1 \\
\hline  
  
\end{tabular}
\end{center}
\end{table*}

\subsubsection{Cool stars}\label{feh}

For the cool stars we derived stellar atmospheric parameters from their FEROS spectra. We took 1D LTE atmospheric models, specifically MARCS spherical models with $1\,$M$_{\sun}$ \citep{marcs}. 
We generated a grid of synthetic spectra by using the radiative transfer code {\scshape spectrum} \citep{graco94}. Although MARCS models are spherical, {\scshape spectrum} treats them as 
if they were plane-parallel. Therefore, the plane parallel transfer treatment might produce a small inconsistency in the calculations of synthetic 
spectra based on MARCS atmospheric models. However, the study of \citet{hei06} concluded that any difference introduced by the spherical models in a plane-parallel transport scheme is small. 
The microturbulent velocity ($\xi$) was fixed according to the calibration given in \citet{dutrafe16}. Effective temperature (T$_{\textrm{eff}}$) ranges from 3\,300~K to 7\,500~K with a 
step of 100~K up to 4\,000~K and 250~K until 7\,500~K, whereas surface gravity ($\log\,g$) varies from $-0.5$ to 3.5~dex in 0.5~dex steps. Finally, in the case of metallicity (using [Fe/H] 
as a proxy), the grid covers from $-1.5$ to $+1.0$~dex in~0.25 dex steps.

We employed a methodology to derive stellar atmospheric parameters based on the iron linelist provided by \citet{Ge13}, since Fe lines are numerous as well as very sensitive in cool stars. 
The linelist contains $\sim$230 features for \ion{Fe}{i} and $\sim$55 for \ion{Fe}{ii}. Their atomic parameters were taken from the VALD database\footnote{\url{http://vald.astro.uu.se/}} 
\citep{pis95,kup2000}. For the Van der Waals damping data, we took the values given by the Anstee, Barklem, and O'Mara theory, when available in VALD \citep[see ][]{bar00}. 

As a starting point, we employed an updated version of the {\scshape stepar} code \citep{hugo18}, adapted to the present problem, that uses stellar synthesis instead of an $EW$ method. 
As optimization method we used the Metropolis--Hastings algorithm. Our method generates simultaneously 48 Markov-chains of 750 points each one. It then performs a Bayesian parameter 
estimation by employing an implementation of Goodman $\&$ Weare's Affine Invariant Markov chain Monte Carlo Ensemble sampler \citep{mcmc}. 
As objective function we used a $\chi^2$ in order to fit the selected iron lines. We fixed the stellar rotation to that value previously derived, and the instrumental broadening to 
the resolution of the FEROS spectrograph. We left the macroturbulent broadening as a free parameter to absorb any residual broadening. 

In the case of the M-(super)giant, S1002, since its spectrum is dominated by molecular bands which erode the continuum weakening or erasing other spectral features such as most iron lines, we were
forced to modify our methodology. We followed the procedure described in \citet{anib07}, focusing on the region 6670--6730\,\AA{} where the TiO bands at 6681 and 6714\,\AA{} are clearly present. 
These bands are very useful since their depth is very sensitive to temperature \citep{anib07}. We used the same grid of MARCS synthentic spectra but using {\scshape turbospectrum} \citep{turbo} 
as a transfer code. 
For this star we computed its stellar parameters based on a $\chi^2$-minimization code implemented by us.

Results (i.e. effective temperature, surface gravity, macroturbulent velocity and iron abundance) are displayed in Table~\ref{Par_2345}.
From the analysis of the six GK (super)giants, we derive a subsolar metallicity for the cluster. The weighted mean (using the variances as weights) is [Fe/H]$=-0.28\pm0.07$, quoting the 
uncertainty in terms of the standard deviation.
Three of these cool objects, stars 14, 50 and 1000, have been observed in three different epochs separated by $\approx$5 years. Spectral variability is not appreciated and 
the parameters derived from individual spectra confirm the consistency of our code. The average difference between parameters from different epochs are $\Delta$\,$T_{\textrm{eff}}$=20\,K, 
$\Delta$\,$\log\,g$=0.09 dex and $\Delta$\,[Fe/H]=0.04 dex, values rather smaller than the intrinsic errors provided by {\scshape StePar}. Once we verified this similarity we listed in 
Table~\ref{Par_2345} the mean parameters for each star, weighting the average by the $S/N$ of the individual spectrum.

\begin{table*}
\caption{Number of FEROS spectra ($N$) for cool stars and average stellar atmospheric parameters derived from them. \label{Par_2345}}
\begin{center}
\begin{tabular}{lccccccccc}   
\hline\hline
Star & Sp T & N & $S/N$ & $v_{\textrm{rad}}$ (km\,s$^{-1}$) & $v \sin\,i$ (km\,s$^{-1}$) &  $\zeta$ (km\,s$^{-1}$) & $T_{\textrm{eff}}$ (K) & $\log\,g$ & [Fe/H]\\
\hline
14  & K3\,Ib       & 3 & 82  & 58.92 $\pm$ 0.12 & 4.9 $\pm$ 1.3 &  4.83 $\pm$ 0.26 & 4\,014 $\pm$ 47 & 0.72 $\pm$ 0.13 & $-$0.33$\pm$ 0.06\\
34$^{*}$  & G6\,II & 2 & 140 & 64.16 $\pm$ 0.13 & 8.1 $\pm$ 1.7 & 10.10 $\pm$ 0.22 & 4\,801 $\pm$ 50 & 1.02 $\pm$ 0.14 & $-$0.19 $\pm$ 0.05\\
43  & K0\,Ib-II    & 2 & 102 & 58.34 $\pm$ 0.08 & 4.9 $\pm$ 1.2 &  4.93 $\pm$ 0.44 & 4\,246 $\pm$ 25 & 1.06 $\pm$ 0.09 & $-$0.29 $\pm$ 0.04\\ 
50  & K4\,Ib       & 3 & 83  & 59.02 $\pm$ 0.38 & 5.6 $\pm$ 1.2 &  3.85 $\pm$ 1.12 & 3\,948 $\pm$ 45 & 0.69 $\pm$ 0.14 & $-$0.34 $\pm$ 0.07\\
60  & K2\,Ib-II    & 2 & 92  & 57.85 $\pm$ 0.02 & 5.5 $\pm$ 0.8 &  5.68 $\pm$ 0.17 & 4\,183 $\pm$ 52 & 0.96 $\pm$ 0.09 & $-$0.28 $\pm$ 0.07\\
1000& K4-5\,Ib     & 2 & 73  & 59.03 $\pm$ 0.22 & 5.9 $\pm$ 1.8 &  3.64 $\pm$ 0.71 & 3\,873 $\pm$ 48 & 0.66 $\pm$ 0.17 & $-$0.25 $\pm$ 0.09\\
1002& M6\,II       & 1 & 60  & 59.69 $\pm$ 0.17 &   ---         &    ---           & 3\,000 $\pm$ 100&  1.5 $\pm$ 0.5  &    0.10 $\pm$ 0.25 \\
\hline

\end{tabular}
\end{center}
\begin{list}{}{}
\item[]$^{*}$ This star exhibits a composite spectrum, whose companion is a blue star (B:).
  \end{list}
\end{table*}

\subsection{Chemical abundances}

\subsubsection{Blue stars}

We tried to derive chemical abundances for hot stars, where the main optical transitions of the chemical elements included in the stellar
atmosphere grid were observed \citep[see][]{Ca12}. Early A-type stars fall outside the range
of effective temperatures included in the {\scshape fastwind} grid (i.e. $T_{\textrm{eff}}$ <\,10\,000\,K). In those stars, it was not possible to perform the chemical analysis, 
using the {\scshape fastwind} grid built in this study. We only employed high-resolution spectra
from FEROS since the resolution of the IDS/ISIS spectra are not the required for this sort of analysis. 
Finally, we only obtained abundances for two B-type stars, namely S28 and LS\,193.

Once we fixed the stellar parameters, we used them to compute tailored models for each star by varying in steps of 0.2 dex the abundances of the chemical elements under study.
In these stars only the line of \ion{Mg}{ii} at 4481\,\AA{} is strong enough to perform a good analysis. \ion{Si}{ii} and \ion{C}{ii} features are also visible but not as 
clear as \ion{Mg}{ii}. The chemical analysis is performed by employing a $\chi^2$ algorithm but a subsequent visual inspection is required to avoid misleading results due to 
weak or blended lines. The technique, as well as the lines considered in the analysis, is described in detail by \citet{Ca12}. Results are
listed in Table~\ref{Abund_cal_2345}. 

\begin{table}
\caption{Chemical abundances, relative to solar abundances by \citet{Gre07}, measured on the blue stars observed with FEROS. \label{Abund_cal_2345}}
\begin{center}
\begin{tabular}{lccc}   
\hline\hline
Star & [Si/H] & [Mg/H] & [C/H] \\
\hline
28     & $-$0.50 $\pm$ 0.67 & $-$0.70 $\pm$ 0.53 & $-$0.30 $\pm$ 0.51 \\ 
LS\,193& $-$0.50 $\pm$ 0.39 & $-$0.90 $\pm$ 0.36 & $-$0.30 $\pm$ 0.25 \\
\hline
  
\end{tabular}
\label{Abund_cal_2345}
\end{center}
\end{table}

\subsubsection{Cool stars}

In the case of the cool stars we computed chemical abundances for the six GK-(super)giants found in the cluster. We employed the same methodology as in the study of NGC\,3105 \citep{3105}. 
We used a method based on $EW$s measured in a semi-automatic fashion using {\scshape tame} \citep{kan12}
for Na, Mg, Si, Ca, Ti, Ni, Y, and Ba. We also measured $EW$s by hand for two special and delicate cases, namely oxygen and lithium using the {\scshape iraf} {\scshape splot} task.
For lithium, we employed a classical analysis using the 6707.8\,\AA{} line, taking into account the nearby \ion{Fe}{i} line at 6707.4\,\AA{}. We measured the $EW$ by hand (in m\AA{}), using the 
{\scshape iraf} {\scshape splot} task. We use the standard notation, where $A($Li$)= \log\,$[n(Li)/n(H)]\,+\,12. In the case of oxygen, we used the [\ion{O}{i}] 6300\,\AA{} line. This oxygen line 
is blended with a \ion{Ni}{i} feature; we corrected the $EW$s accordingly by using the methodology and line atomic parameters described in \citet{bel15}. Finally we also derive rubidium 
abundances using stellar synthesis for the 7800\,\AA{} \ion{Rb}{i} line, following the methodology in \citet{dor13}.

Table~\ref{Abund_cool_2345} summarises the abundances found for each star together with their errors. We estimate the cluster average by using a weighted arithmetic mean (employing the variances
as weights). We computed conservative errors, since the associated uncertainty is the combination in quadrature of the individual typical error and the star-to-star dispersion. 

\begin{table*}
\caption{Chemical abundances, relative to solar abundances by \citet{Gre07}, measured on the cool stars.}
\begin{center}
\begin{tabular}{lcccccc}   
\hline\hline
Star & [O/H] &[Na/H] & [Mg/H] & [Si/H] & [Ca/H] & [Ti/H]\\
\hline
14      & $-$0.21 $\pm$ 0.11 & $-$0.19 $\pm$ 0.36 & $-$0.14 $\pm$ 0.06 &    0.00 $\pm$ 0.17 & $-$0.05 $\pm$ 0.20 & $-$0.08 $\pm$ 0.18 \\
34      & $-$0.18 $\pm$ 0.09 &    0.08 $\pm$ 0.22 & $-$0.10 $\pm$ 0.05 &    0.07 $\pm$ 0.12 &    0.11 $\pm$ 0.13 & $-$0.15 $\pm$ 0.14 \\
43      &    0.02 $\pm$ 0.07 & $-$0.12 $\pm$ 0.18 & $-$0.03 $\pm$ 0.07 &    0.08 $\pm$ 0.10 & $-$0.08 $\pm$ 0.14 & $-$0.19 $\pm$ 0.16 \\
50      & $-$0.22 $\pm$ 0.09 & $-$0.10 $\pm$ 0.37 & $-$0.05 $\pm$ 0.06 & $-$0.04 $\pm$ 0.14 &    0.07 $\pm$ 0.17 &    0.09 $\pm$ 0.15 \\
60      & $-$0.04 $\pm$ 0.09 &    0.00 $\pm$ 0.27 & $-$0.02 $\pm$ 0.05 &    0.04 $\pm$ 0.17 & $-$0.02 $\pm$ 0.17 & $-$0.05 $\pm$ 0.16 \\
1000    & $-$0.19 $\pm$ 0.11 & $-$0.15 $\pm$ 0.41 & $-$0.11 $\pm$ 0.11 &    0.01 $\pm$ 0.15 &    0.13 $\pm$ 0.15 &    0.17 $\pm$ 0.22 \\
\hline
Mean    & $-$0.12 $\pm$ 0.09 & $-$0.06 $\pm$ 0.31 & $-$0.07 $\pm$ 0.07 &    0.04 $\pm$ 0.14 &    0.03 $\pm$ 0.16 & $-$0.15 $\pm$ 0.17 \\
&                     &                    &                    &                    &                    & \\

\hline\hline
Star & [Ni/H] & [Rb/H] & [Y/H] & [Ba/H] & $EW$(Li) & A(Li)\\
\hline
14      & $-$0.12 $\pm$ 0.14 &   $-$0.35 $\pm$ 0.02  & $-$0.16 $\pm$ 0.46 & 0.34 $\pm$ 0.10 & 237.1 & 1.60 $\pm$ 0.15 \\
34      & $-$0.25 $\pm$ 0.08 &   ---                 & $-$0.13 $\pm$ 0.21 & ---             & ---   & ---             \\
43      & $-$0.15 $\pm$ 0.08 &   ---                 & $-$0.28 $\pm$ 0.26 & 0.38 $\pm$ 0.07 & ---   & ---             \\
50      & $-$0.04 $\pm$ 0.15 &   $-$0.30 $\pm$ 0.02  & $-$0.26 $\pm$ 0.17 & 0.38 $\pm$ 0.10 & 286.7 & 2.11 $\pm$ 0.11 \\
60      & $-$0.09 $\pm$ 0.14 &   $-$0.33 $\pm$ 0.04  & $-$0.11 $\pm$ 0.42 & 0.37 $\pm$ 0.10 & 81.2  & <\,0.58         \\
1000    &    0.00 $\pm$ 0.18 &   $-$0.30 $\pm$ 0.01  & $-$0.04 $\pm$ 0.29 & 0.43 $\pm$ 0.09 & 252.6 & 1.65 $\pm$ 0.04 \\
\hline
Mean    & $-$0.15 $\pm$ 0.13 &  $-$0.31 $\pm$ 0.02 & $-$0.19 $\pm$ 0.32 & 0.38 $\pm$ 0.09 & & \\
\hline

\end{tabular}

\label{Abund_cool_2345}
\end{center}
\end{table*}

\section{Discussion}

In this work we performed a comprehensive study of the open cluster NGC\,2345. The combination of Str\"{o}mgren photometry, $Gaia$ data and spectroscopy at different 
resolutions makes our analysis, by far, the most robust of all those carried out to date.

\subsection{Cluster parameters}

Cluster parameters such as reddening, size, distance and age obtained for NGC\,2345 by different authors are displayed in Table~\ref{cum_param} for comparison. As mentioned in Sect.~\ref{intro}
there are two main photometric works about this cluster. The first one \citep{mof74} employed $UBV$ photoelectric and photographic photometry whereas the second one \citep{carr15}
performed for the first time modern CCD photometry. We place the cluster roughly at an equidistant point between previous estimates. The age obtained in this work is slightly younger, although compatible 
within the errors, than that given by \citep{carr15}, which assumed a solar metallicity instead of a subsolar value as we did. 

\begin{table*}
\caption{Summary of cluster parameters for NGC\,2345 derived in this work compared to studies found in the literature. The solar reference used for calculating Galactocentric 
distances is $R_{\sun}=8.3$~kpc.}
\begin{center}
\begin{tabular}{lccccc}   
\hline\hline
Reference      & Phot. &  $E(B-V)$  & $d$ (kpc)           & $r_{\textrm{cl}}$\,($\arcmin$)& Age (Ma)\\
\hline     
\citet{mof74}  & Pg+Pe & 0.48--1.16         & 1.75              & 5.25--6.15                & 60      \\ 
\citet{carr15} & CCD   & 0.59\,$\pm$\,0.04  & 3.0\,$\pm$\,0.5   &  3.75                     & 63--70  \\
This work      & CCD   & 0.37--1.17         & 2.5\,$\pm$\,0.2   & 18.7\,$\pm$\,1.2          & 56\,$\pm$\,13\\
\hline
\end{tabular}

\label{cum_param}
\end{center}
\end{table*}

\subsubsection{Supergiants}

The number and type of supergiants contained in a cluster is an important observable to constrain the stellar evolutionary models. 
NGC\,2345 hosts two A- and six K-type supergiants, representing a blue-to-red supergiant ratio, $B/R$=0.33, similar to that observed in NGC\,3105 ($B/R$=0.4).
Both values are approximately half of the expected value, $B/R$=0.7, according to the empirical expression found by \citet{egg02}. 
As discussed in \citet{be51,3105}, A supergiants are mostly observed in young clusters as objects leaving the MS and not as blue-loop objects, against the prediction of
Geneva models \citep{georgy13}. In addition, no yellow supergiants are observed in NGC\,2345 in contrast to Be\,51 \citep{be51}, a coeval cluster with a similar
mass which hosts four yellow supergiants but none blue.

\subsection{The extent of NGC\,2345}

The main difference with respect the other works is found in the cluster size: our value for the radius is more than three times larger than that found by \citet{mof74} and five times in the case of the value estimated by \citet{carr15}.
This result, obtained by using photometry, is consistent with the position of the most distant members confirmed via RV (such as LS\,193 at $\approx$17$'$). 
Additionally, we have identified among the most distant objects observed with AA$\Omega$ some stars (LS\,174, LS\,176, LS\,205 and LS\,229), at distances from 25$'$ up to 50$'$, with RV and astrometric data compatible
with those of the cluster.

\begin{figure}
  \centering         
  \includegraphics[width=\columnwidth]{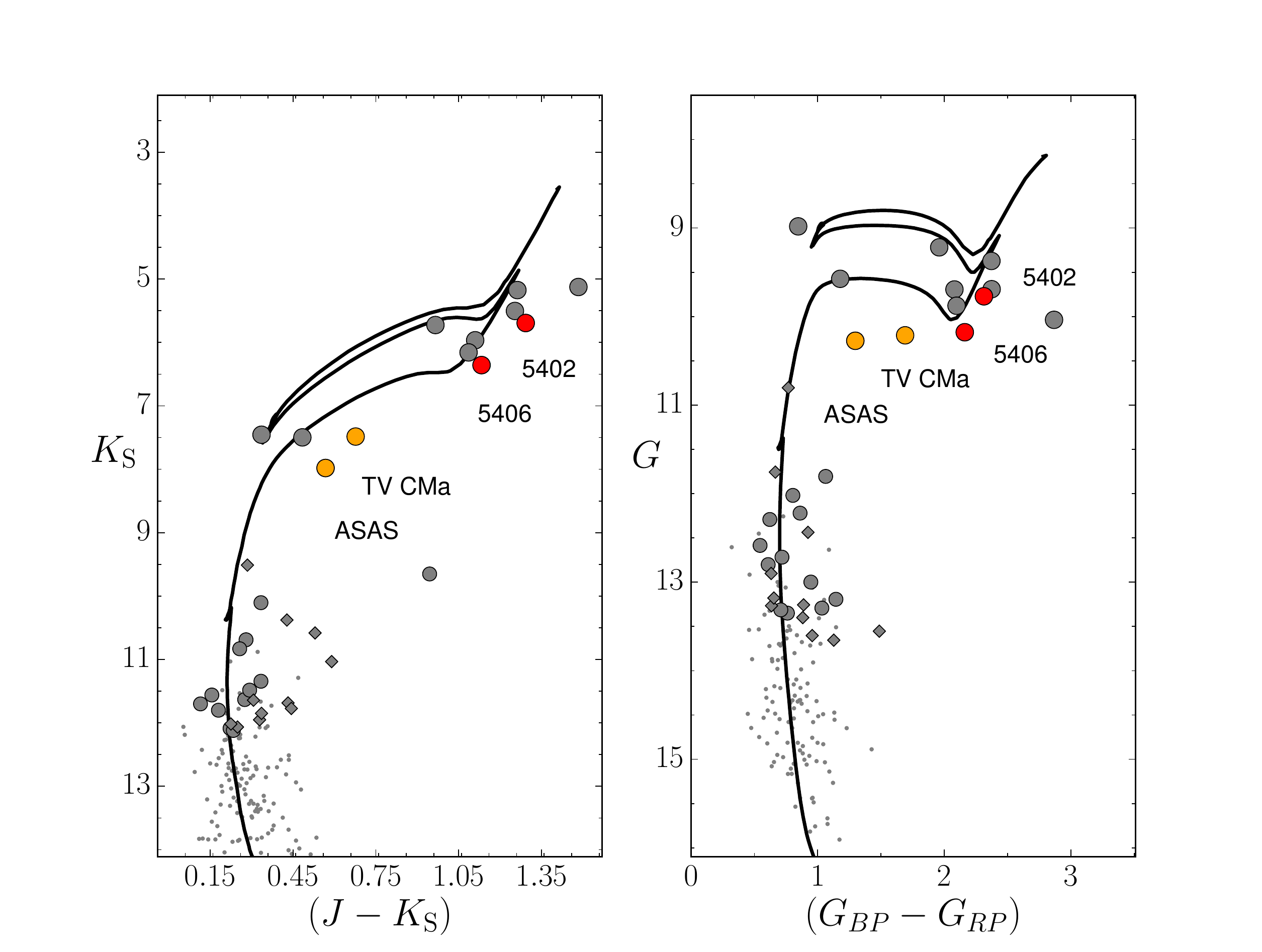}   
  \caption{2MASS (left) and $Gaia$ (right) colour-magnitude diagrams as in the Fig.\ref{isocronas} but highligthing the Cepheids (orange circles) and the likely new 
  red supergiants (red circles) described in the text. For clarity, the names of each star have been shortened.}
  \label{nuevas}
\end{figure}

\citet{An13} evaluated the membership of the Cepheids TV\,CMa and ASAS\,J070911-1217.2, both objects located at a great distance from the nominal cluster centre (38$\arcmin$ and 56$\arcmin$, respectively).
In the first case, TV\,CMa, they discarded its relationship with NGC\,2345 whereas for the second Cepheid they were not able to yield a conclusive result. 
In the light of the new $Gaia$ DR2 data we put our attention back on this issue. The proper motions and the RV (37$\pm$7 km\,s$^{-1}$) are not compatible with the membership for TV\,CMa, 
as they suggested. Instead, ASAS\,J070911-1217.2 shows astrometric values in good agreement with the cluster average. Only the RV (70$\pm$3 km\,s$^{-1}$) is somewhat different from what is expected 
for a member, even taking into account that it is a variable star. In any case, it seems unlikely that this Cepheid is associated to the cluster, since it is very distant and its position on the CMDs falls away from the isochrone. 
(see Figure\,\ref{nuevas}).

In addition, we have identified two new possible red (super)giants among all likely members previously selected (Sect.\ref{membership}). These objects were not covered neither by our photometry nor
spectroscopically observed. The first one, TYC\,5402-1851-1, is located at 22$\arcmin$ from the nominal cluster centre whereas the second one, TYC\,5406-46-1, is 25$\arcmin$ away.
Both stars share the average astrometric parameters of the cluster but their RVs are higher than the mean value (68.6$\pm$0.3 km\,s$^{-1}$ and 77.0$\pm$0.2 km\,s$^{-1}$, respectively).
On the other hand, their positions on the CMDs (in contrast to the Cepheids) are close to those of other evolved stars which could indicate a relationship with the cluster.
Figure\,\ref{nuevas} shows the position of these four evolved stars on the 2MASS and $Gaia$ CMDs.

The detection of all these bright stars in the cluster surroundings, sharing the average values of the cluster, might suggest the existence of a large halo spreading around NGC\,2345, beyond the borders 
estimated by us in this work. These $halo$ $members$ are marked in Fig.\,\ref{2345_grande} to facilitate their location with respect to the cluster. Likewise,
their astrometric data are also displayed in Table\,\ref{tab_mp}.

\subsection{Be stars}

NGC\,2345 is known to host a large number of Be stars. Among the whole sample studied by \citet{be_2345} NGC\,2345 is the cluster with the highest fraction of Be stars. \citet{be_2345, mathew11}
identified for the first time 12 Be stars, representing a Be fraction, i.e. Be/(B+Be), of $\approx$\,26\,$\%$. Instead, from our spectroscopy, among the likely members, we found 12 Be stars, nine of which are in common with them 
(Table~\ref{tab_be}). Without counting the B companion in S34, we also identified 14 normal B-type stars, which implies that almost half of the B stars for which we took spectra show the emission phenomenon.

According to \citet{Po03} classical Be stars are non-supergiant B-type stars that show or had shown Balmer lines in emission at any time. Thus, with the aim of including all the Be stars in NGC\,2345, 
we add to our sample those Be stars previously identified by \citet{be_2345} in case they were likely members. These stars, namely S27, X1 and X2, show astrometric values
compatible with cluster membership (displayed in Table~\ref{tab_mp}) and, in fact, they had already been selected as likely B members (which proves the validity of our photometric analysis).
Commonly, in young open clusters Be stars are concentrated around the MSTO, \citep{Sw05} as observed in NGC\,2345, where they are located in the upper MS,
with spectral types in the range B3--B7 (Table ~\ref{Be_2345}). We found a real fraction of Be stars for NGC\,2345, i.e. considering all likely B members and not only those observed spectroscopically, 
of 10.3\,$\%$.
This value can be considered as a lower limit since we are assuming that there are no more Be stars among the B members selected photometrically. 
Still in this case NGC\,2345 is one of the clusters with the highest Be fraction among those studied by \citet{be_2345}.

The Be phenomenon has been observed in young open clusters hosting B stars, with ages comprised approximately between 10 and 300 Ma. \citet{Me82} suggested that the Be fraction 
decreases with increasing age. In the Galaxy, Be stars in open clusters show a bimodal distribution peaking around spectral types B1--B2  and B7--B8 \citep{Me82} or B1--B2 and B5--B7 \citep{be_2345}. 
In NGC\,2345 Be stars are concentrated around spectral type B5 and 83\,$\%$ are found in the bin B5--B7 (see Table~\ref{Be_2345}), as it has been observed in other clusters. 
However, clusters with high Be fractions, such as NGC\,663 \citep{Pi01}, NGC\,7419 \citep{7419} or NGC\,3105 \citep{3105}, seem to be grouped around ages in the range 15--40 Ma. 
This is in an excellent agreement 
with theoretical considerations that associate the generation of circumstellar discs, responsible for the emission observed, with the rotational velocity distribution expected among early stars \citep{Gra16}.
NGC\,2345 shows a high Be fraction at an atypical age \citep[see the figure 8 in][]{be_2345}.

Finally, we have to keep in mind that not only the age is key to explain the Be phenomenon but also other factors such as the geometry of the cluster and the metallicity.
By the one hand, the shape of the cluster is related to the initial angular momentum with which the cluster originated \citep{kel99}. Thus, a high eccentricity would be the consequence 
of a rapid rotation rate and, therefore, a high Be fraction would be expected. By the other hand, low-metallicity environments are known to favor the Be phenomenon \citep{mae00,Mar07} as we argued in more 
detail when studied NGC\,3105 \citep[see the discussion in][]{3105}, a younger metal-poor cluster with a comparable Be fraction.

\begin{table}
\caption{Fraction of Be stars in NGC\,2345 among those members for which we have spectra.}
\begin{center}
\begin{tabular}{lccc}   

\hline\hline
Sp T& B & Be & Fraction ($\%$)\\
\hline
B3 & 8 & 1 & 11.1\\
B4 & 2 & 1 & 33.3\\
B5 & 2 & 7 & 77.8\\
B6 & 2 & 0 & 0.0\\
B7 & 0 & 3 & 100.0\\
\hline
Total & 14 & 12 & 46.2\\
\hline
\end{tabular}

\label{Be_2345}
\end{center}
\end{table}

\subsection{Atmospheric stellar parameters}

As mentioned above (Sect.~\ref{intro}) we found in the literature two spectroscopic studies carried out on NGC\,2345 before this work.
\citet{reddy16} derived stellar parameters and abundances only for three of the cool stars, namely 34, 43 and 60. In Table~\ref{Comp_par16}
we compare them with the values obtained in this work. For stars 34 and 43 our results are compatible within the incertainties whereas for S60 all our values are lower than theirs.
More recently, \citet{hol19} carried out the spectral analysis of the five red (super)giants. We reanalysed the same spectra following our procedure. 
Both set of values are compatible within the errors. Only the surface gravities for S43 are significantly different (see Table~\ref{Comp_par19}).
Regarding metallicity, our values are fully \citep{hol19} or marginally compatible \citep{reddy16} with previous works while our microturbulent velocities
are somewhat smaller (on average 0.6--0.7 km\,s$^{-1}$).

\begin{table*}
\caption{Comparison of temperature, surface gravity, microturbulent velocity and iron abundance with those of \citet{reddy16}. \label{Comp_par16}}
\begin{center}
\begin{tabular}{l|cccc|cccc}   
\hline\hline
\multirow{2}{*}{Star} & \multicolumn{4}{c|}{Reddy16$^{*}$} & \multicolumn{4}{c}{This work}\\
   & $T_{\textrm{eff}}$ (K)& $\log\, g$ &  $\xi$ (km\,s$^{-1}$) & [Fe/H]$^{**}$ & $T_{\textrm{eff}}$ (K)& $\log\, g$ &$\xi$ (km\,s$^{-1}$) & [Fe/H]\\
\hline
34  & 4\,850  & 0.85 & 3.23 & $-0.21\pm0.04$ & 4\,801 $\pm$ 50 & 1.02 $\pm$ 0.14 & 2.40 & $-0.19\pm0.05$ \\
43  & 4\,300  & 1.20 & 2.21 & $-0.19\pm0.04$ & 4\,246 $\pm$ 25 & 1.06 $\pm$ 0.09 & 1.72 & $-0.29\pm0.04$ \\
60  & 4\,300  & 1.60 & 2.21 & $-0.13\pm0.05$ & 4\,183 $\pm$ 52 & 0.96 $\pm$ 0.09 & 1.73 & $-0.28\pm0.07$ \\
\hline

\end{tabular}
\end{center}
\begin{list}{}{}
\item[]$^{*}$ Typical uncertainties are $\Delta\,T_{\textrm{eff}}$\,=\,75\,K, $\Delta\,\log\, g$\,=\,0.25\,dex and $\Delta\,\xi$\,=\,0.20\,km\,s$^{-1}$.
\item[]$^{**}$ The iron abundance is rescaled to the solar reference used in this work \citep{Gre07}.
\end{list}
\end{table*}

\begin{table*}
\caption{Comparison of temperature, surface gravity, microturbulent velocity and iron abundance with those of \citet{hol19}. \label{Comp_par19}}
\begin{center}
\begin{tabular}{l|cccc|cccc}   
\hline\hline
\multirow{2}{*}{Star} & \multicolumn{4}{c|}{Holanda19$^{*}$} & \multicolumn{4}{c}{This work}\\
   & $T_{\textrm{eff}}$ (K)& $\log\, g$ &  $\xi$ (km\,s$^{-1}$) & [Fe/H]$^{**}$ & $T_{\textrm{eff}}$ (K)& $\log\, g$ &$\xi$ (km\,s$^{-1}$) & [Fe/H]\\
\hline
14  & 4\,150  & 1.10 & 2.36 & $-$0.29 & 4\,024 $\pm$ 52 & 0.77 $\pm$ 0.14 & 1.71 & $-0.31\pm0.07$ \\
34  & 4\,850  & 1.10 & 3.00 & $-$0.20 & 4\,810 $\pm$ 45 & 1.08 $\pm$ 0.09 & 2.35 & $-0.19\pm0.05$ \\
43  & 4\,350  & 1.60 & 2.45 & $-$0.27 & 4\,251 $\pm$ 21 & 1.10 $\pm$ 0.06 & 1.69 & $-0.31\pm0.03$ \\
50  & 4\,000  & 0.70 & 2.31 & $-$0.32 & 3\,948 $\pm$ 34 & 0.70 $\pm$ 0.10 & 1.67 & $-0.34\pm0.05$ \\
60  & 4\,020  & 1.03 & 2.54 & $-$0.34 & 4\,200 $\pm$ 45 & 1.00 $\pm$ 0.13 & 1.72 & $-0.27\pm0.06$ \\
\hline

\end{tabular}
\end{center}
\begin{list}{}{}
\item[]$^{*}$ Typical uncertainties are $\Delta\,T_{\textrm{eff}}$\,=\,70--100\,K, $\Delta\,\log\, g$\,=\,0.1--0.2\,dex, $\Delta\,\xi$\,=\,0.20\,km\,s$^{-1}$
and $\Delta$\,[Fe/H]\,=\,0.1\,dex.
\item[]$^{**}$ The iron abundance, is rescaled to the solar reference used in this work \citep{Gre07}.
\end{list}
\end{table*}

\begin{figure} 
  \centering         
  \includegraphics[width=\columnwidth]{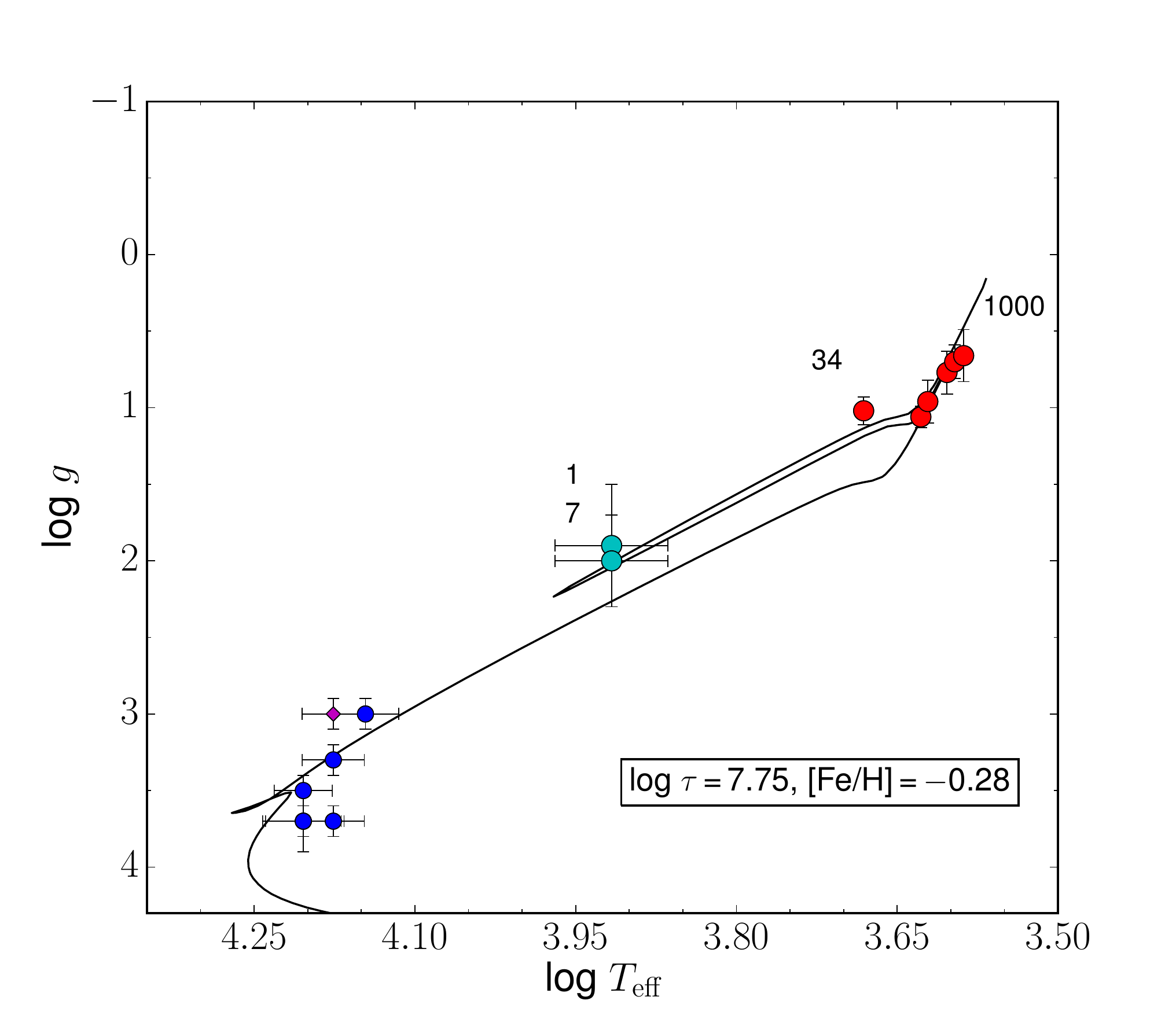}   
  \caption{Kiel diagram for likely members. Colours and symbols are the same as those in Fig.~\ref{c1_m1}. 
  The best-fitting PARSEC isochrone is the black line.}
  \label{pHR_2345}
\end{figure}

Finally, we plot the log\,$g$\,/\,$T_{\textrm{eff}}$ (Kiel diagram) for all cluster members for which we have stellar parameters (Fig.~\ref{pHR_2345}). On this diagram we 
also draw the best-fitting isochrone (solid line). The position of the stars matches pretty well the isochrone (within the errors), especially for the evolved stars.
The location of S1000 on this diagram, showing an excellent agreement with the other K (super)giants, suggests that its anomalous position in the photometric diagrams
could be related to its intrinsic reddening, somewhat higher from the rest of red (super)giants.
Only star 34, which is a binary, is slightly displaced from the expected evolutionary locus. This suggests that binarity may play a role in 
determining its evolution towards the cool side of the diagram.
It is noteworthy the excelent agreement between spectral results (stellar parameters) and cluster parameters inferred photometrically such as the distance and age (which 
are reflected in the choice of the isochrone). This demonstrates the reliability of our results. 

\subsubsection{Low metallicity}

We obtained a metal-poor composition for NGC\,2345 of [Fe/H]$=-0.28\pm0.07$. This value, computed from the six red giants, is compatible with previous estimations: $-0.28\pm0.05$ \citep{hol19} 
and $-0.17\pm0.06$ \citep{reddy16}, which used five and three stars, respectively. For a proper comparison both values have been rescaled to our solar reference \citep{Gre07}.

\begin{figure}  
  \centering         

  \includegraphics[width=\columnwidth]{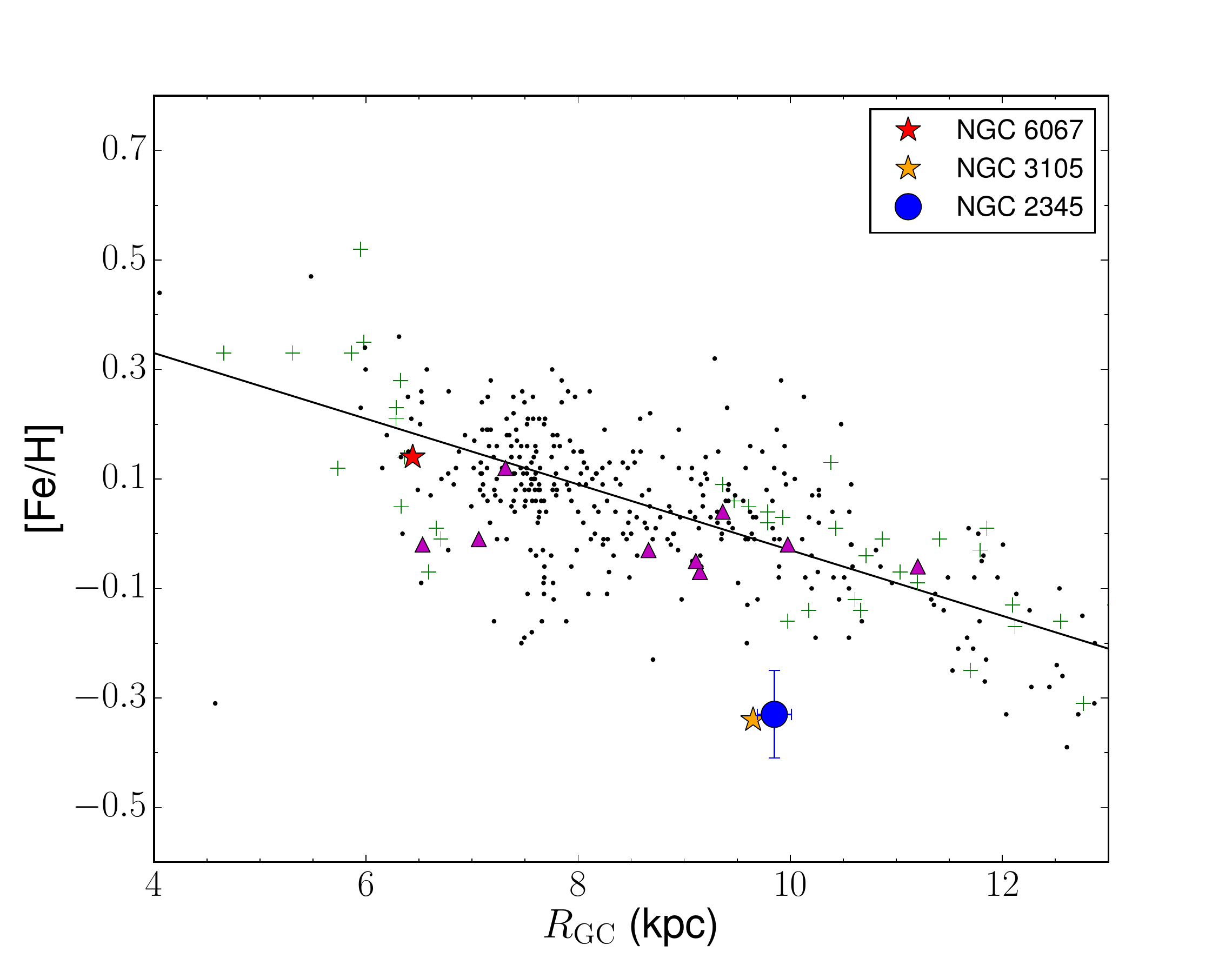}  
  \caption{Iron abundance gradient in the Milky Way found by \citet{Ge13,Ge14}. The black line is the Galactic gradient ($-$0.06~dex/kpc), green crosses are Cepheids studied in those papers,
  whereas black dots show data for other Cepheids from literature used by these authors. Magenta triangles represent the young open clusters ($<500\:$Ma) in the sample compiled by \citet{net16}. Finally, the 
  blue circle is NGC~2345. NGC~6067 and NGC~3105, other young clusters analysed by our group with the same techniques have been added for comparison. NGC~2345, together with NGC~3105, falls well below the metallicity
  typically observed at its Galactocentric distance. All the values shown in this plot are rescaled to \citet{Ge14}, i.e. R$_{\sun}$=7.95 kpc and A(Fe)=7.50.} 
  \label{gradient_2345} 
\end{figure}

In order to put in context our result, we place the cluster on the Galactic gradient, taking as a reference the work by \citet{Ge13,Ge14}. With the aim of estimating the radial distribution of metallicity they
used Cepheids, since they are a young population ($\tau\approx20$\,--\,400~Ma) tracing present-day abundances. In this way, Cepheids offer an appropriate comparison for young clusters such as NGC\,2345. 
In Fig.~\ref{gradient_2345}, we show this gradient with the position of our cluster overplotted as well as a sample of open clusters studied by \citet{net16} for further comparison. 
Among them, we take young clusters (i.e. age below 500 Ma) whose metallicity is derived from, on average, at least three high-quality spectra \citep{hei14}. 
The position of NGC\,2345, very close to NGC\,3105, highlights the low metallicity of both clusters and illustrates the wide range of metallicities found in the outer region of the Galaxy.
These two clusters, at a galactocentric distances around 10\,kpc, show a metallicity comparable to that of the LMC.
\citet{6067} suggested that metallicity inhibits the formation of Cepheids. Nevertheless, the absence of Cepheids in low-metallicity environments such as NGC\,2345 and NGC\,3105 casts doubts on that possibility
since both are massive enough clusters to serve as stellar evolution testbeds.

\subsection{Stellar chemical abundances}

\begin{figure} 
  \centering         
  \includegraphics[width=\columnwidth]{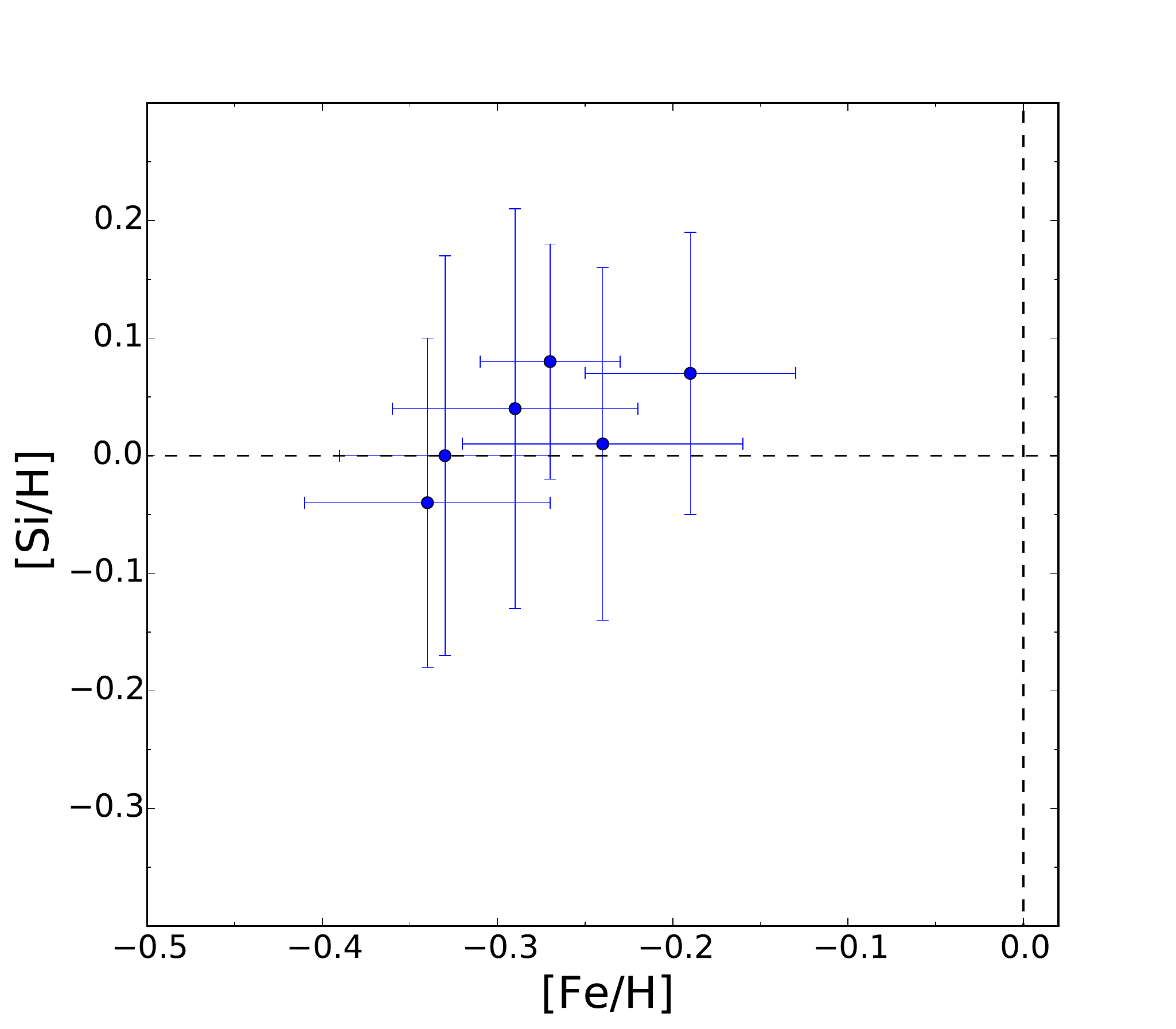}   
  \caption{Abundance ratios [Si/H] vs [Fe/H]. All the stars have the same composition (within the errors), indicating chemical homogeneity. As a reference, the dashed lines show the solar values.}
  \label{disp_2345}
\end{figure}

NGC\,2345 seems to be chemically homogeneous since for every element, abundances derived from different (cool) stars are comparable within the errors and all they are grouped around the cluster average.
In Fig.~\ref{disp_2345} we show the representative star-to-star scatter by plotting the abundance of Si for all the objects in our sample.
We also derived some abundances from hot stars, but with large uncertainties, that yield a metal-poor cluster. From the analysis of both groups of stars, hot and cool, we obtained abundances marginally compatible, within
the errors, for the elements in common (Si and Mg).

We compare the chemical abundances obtained in this work with those previously published. 
By the one hand, \citet{hol19} derived abundances for 19 elements, nine of which are in common with those studied by us. By the other hand, \citet{reddy16} computed them for 23 species, having
in common with us ten. Both sets of abundances as well as ours are displayed in terms of the ratios [X/Fe] in Table\,\ref{comp_abund}.  
We computed compatible abundances, within the errors, for almost all the elements in common, with the exception of Ca and Ba. For these elements our values are up to 0.5 dex higher.

\begin{table*}
\caption{Comparison of the mean abundances ratios, relative to solar abundances by \citet{Gre07}, obtained for NGC\,2345 so far.}
\begin{center}
\begin{tabular}{l|c|c|c}   
\hline\hline
Ratio & Reddy16 & Holanda19 & This work \\
\hline
$[$Na/Fe$]$ & $+$0.21 $\pm$ 0.07 & $+$0.29 $\pm$ 0.12 & $+$0.22 $\pm$ 0.32 \\    
$[$Mg/Fe$]$ & $-$0.02 $\pm$ 0.02 & $+$0.20 $\pm$ 0.08 & $+$0.21 $\pm$ 0.10 \\
$[$Si/Fe$]$ & $+$0.20 $\pm$ 0.05 & $+$0.24 $\pm$ 0.08 & $+$0.32 $\pm$ 0.16 \\
$[$Ca/Fe$]$ & $-$0.25 $\pm$ 0.10 & $-$0.02 $\pm$ 0.03 & $+$0.31 $\pm$ 0.17 \\
$[$Ti/Fe$]$ & $-$0.16 $\pm$ 0.08 & $+$0.03 $\pm$ 0.07 & $+$0.13 $\pm$ 0.18 \\
$[$Fe/H$]$  & $-$0.15 $\pm$ 0.06 & $-$0.33 $\pm$ 0.05 & $-$0.28 $\pm$ 0.07 \\
$[$Ni/Fe$]$ & $-$0.15 $\pm$ 0.06 & $-$0.12 $\pm$ 0.01 & $+$0.13 $\pm$ 0.15 \\
$[$Y/Fe$]$  & $+$0.05 $\pm$ 0.05 & $+$0.19 $\pm$ 0.06 & $+$0.09 $\pm$ 0.33 \\
$[$Ba/Fe$]$ & $+$0.14 $\pm$ 0.04 &                    & $+$0.66 $\pm$ 0.11 \\
\hline

\end{tabular}

\label{comp_abund}
\end{center}
\end{table*}

\begin{figure*}  
  \centering         
  \includegraphics[width=19cm]{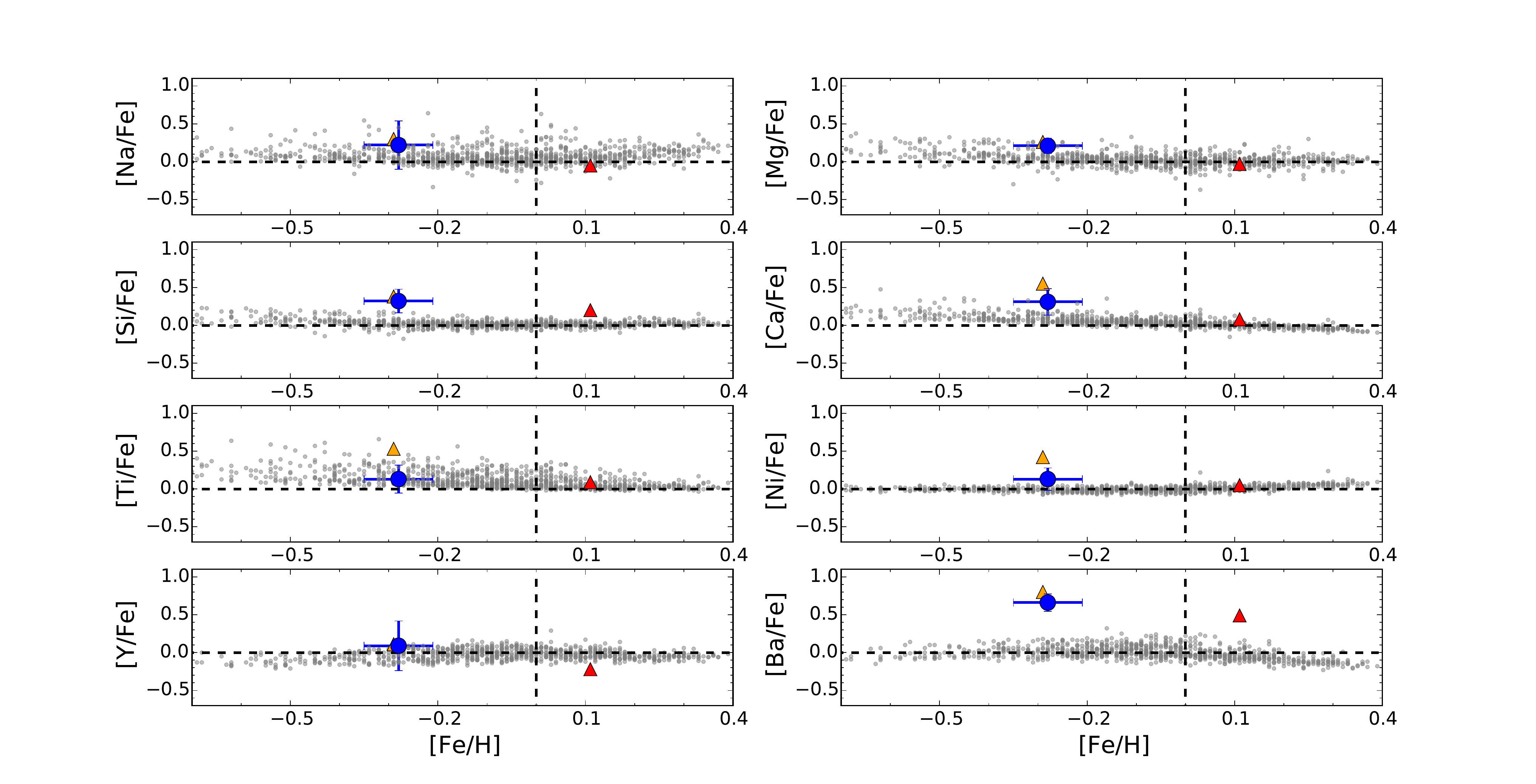}   
  \caption{Abundance ratios [X/Fe] versus [Fe/H]. The grey dots represent the galactic trends for the thin disc \citep{vardan,elisa17}. NGC~6067 and NGC~3105 are drawn with triangles (orange and red, respectively)  
  whereas NGC\,2345 is the blue circle. Clusters are represented by their mean values. The dashed lines indicates the solar value.} 
  \label{trends_2345} 
\end{figure*}

Finally, we also compare our abundances with the Galactic trends for the thin disc (Fig.~\ref{trends_2345}). We plot abundance ratios [X/Fe] vs [Fe/H] obtained by \citet{vardan} for Na, Mg, Si, Ca, Ti and Ni and by \citet{elisa17} for Y and Ba in the workframe 
of the HARPS GTO planet search program. The chemical composition of NGC\,2345 is compatible, within the errors, with the Galactic trends observed in the thin disc of the Galaxy.
We derived a roughly solar [Y/Fe] against a supersolar [Ba/Fe], which is in good agreement with the dependence on age and Galactic location found by \citet{Mi13} by comparing the
abundances of Y and Ba in different open clusters. 
Remarkably, we find a strong over-abundance of Barium ([Ba/Fe]=$+$0.8). Several young open clusters have been observed to have Ba abundances higher than those predicted by standard theoretical models.
To explain this enrichment of Ba, \citet{dor09} suggested an enhanced ``$s$-process''. Our result supports this idea (see their figure~2).

\subsubsection{Lithium-rich stars}

\begin{figure} 
  \centering         
  \includegraphics[width=\columnwidth]{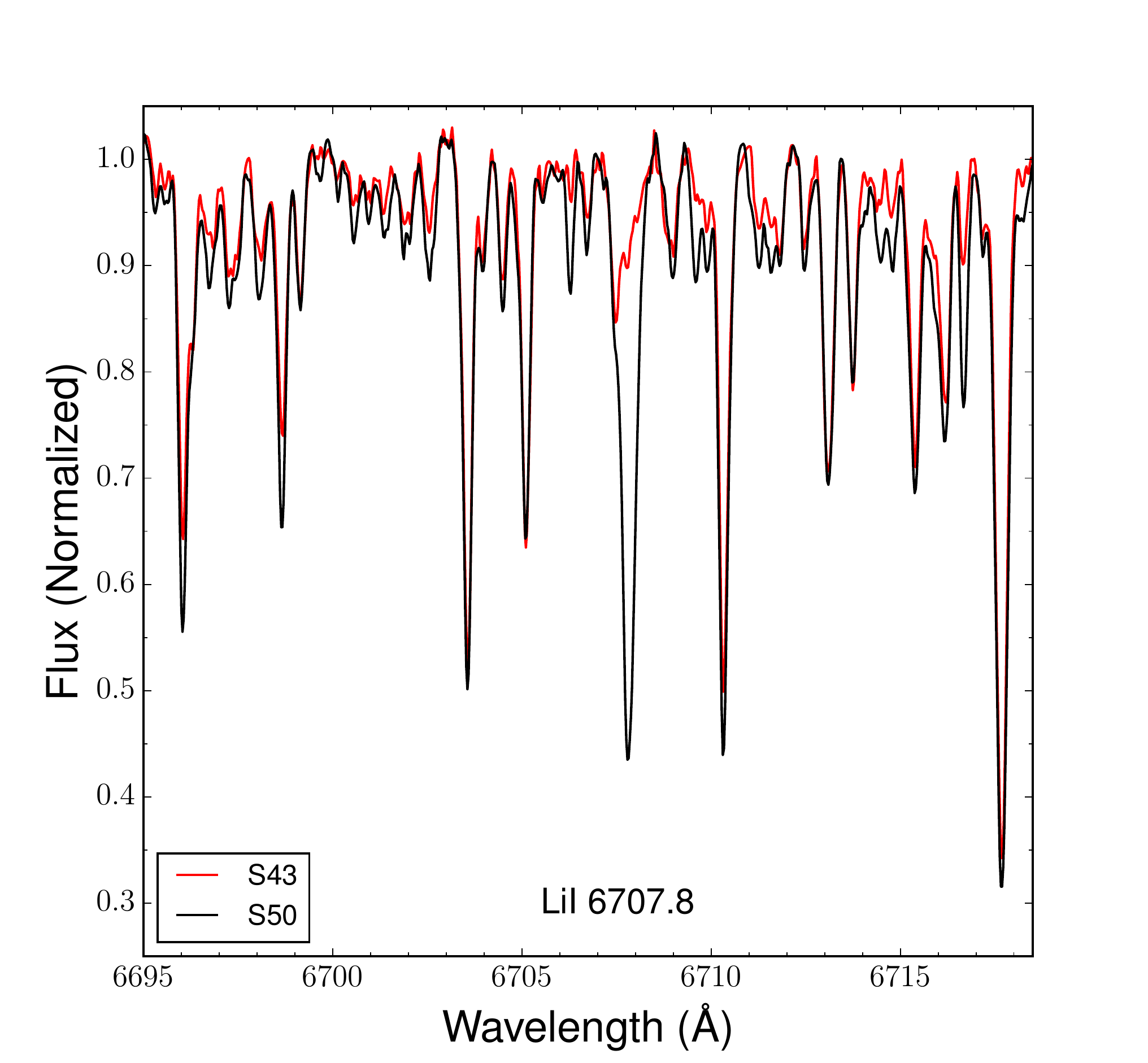}  
  \caption{Spectra around the Li line at 6707\,\AA. Li-rich giant, star 50, is represented versus star 43, a Li-normal giant with similar stellar parameters.}
  \label{litio_2345}
\end{figure}

Canonical models \citep{ib67,ib67_2,sod93} predict Li depletion once stars reach the red giant branch as a direct consequence of the first dredge-up. According
to this, we do not expect to find in red giants abundances of Li above 1.5 dex \citep{Ch00}. In good agreement with this scenario, Li is not found, or just in a 
small amount in stars 34, 43 and 60. Nevertheless, as in the case of NGC\,6067 \citep{6067}, we find three Li-rich stars: S14, S50 and S1000. For stars S14 and S1000, the abundance of Li 
(1.67$\pm$0.13 and 1.65$\pm$0.04, respectively.) could be marginally compatible with the canonical scenario, but not for S50. It exhibits a high lithium abundance, A(Li)=2.14, a value 12 times greater 
than solar. This fact is clearly evidenced when comparing the spectrum of this Li-rich star with that of a star in which Li is not detected (see Fig.~\ref{litio_2345}). These stars,
as S276 in NGC\,6067 \citep{6067}, are the coolest in the cluster and none of them exhibits a noticeable abundance of Rb, which discards the remote possibility that they were AGBs. 

\citet{reddy16} did not compute Li abundances for any star but \citet{hol19} gave them for the five giants. Their values are displayed together with ours in Table\,~\ref{comp_li}.
When we compare both sets of abundances we note that their values are significantly smaller than ours, specially for S50. 

\begin{table}
\caption{Abundance of Li for five red giants in NGC\,2345}
\begin{center}
\begin{tabular}{l|c|c}   
\hline\hline
Star & Holanda19$^{*}$ & This work \\
\hline
14 & 0.95  &  1.67 $\pm$ 0.09  \\
34 & 0.11  &  ---              \\  
43 & 0.31  &  ---              \\ 
50 & 0.81  &  2.14 $\pm$ 0.03  \\
60 & 0.19  &  $<$0.53          \\
\hline

\end{tabular}

\label{comp_li}
\end{center}
\begin{list}{}{}
\item[]$^{*}$ Typical uncertainties are $\pm$\,0.1\,dex.
  \end{list}
\end{table}

These Li-rich giants, a fraction $\sim$1\,$\%$ of all the giants analysed, have been found both in isolation \citep{Br89} and within open clusters \citep{elisa16, 6067}. 
Different scenarios have been proposed to explain this overabundace. The \citet{CFM} mechanism (CFM) can produce Li via the hot botton burning \citep{hbb} in 
intermediate-mass stars during the AGB phase as well as during the RGB phase in low-mass stars \citep{sac99}, although it requires an 
extra mixing process. The engulfment of a sub-stellar companion \citep{Si99,Ag16} or the enrichment of the interstellar medium after the explosion of a SN \citep{Woo95} could 
also make sense of the large amount of Li found. 

As the cluster chemical composition (and S50 itself) is not above the galactic average, the SN explosion scenario
can be discarded. Given its very low abundance of $s$-process elements, such as Rb, and its position on the CMD, S50 should not be 
an AGB star, and so the CFM is also rejected. After discarding all other possibilities, the engulfment of a planet or a brown dwarf by the star could be a plausible explanation,
but our data cannot confirm or discard it. To shed some light on this issue, we should determine the $^6$Li/$^7$Li ratio, for which a spectrum with rather higher resolution is needed.

\section{Conclusions}

We performed an in-depth analysis of NGC\,2345, the most complete to date. We provided the largest set of spectra, most of them for classification purposes, which
combined with photometry and astrometry has allowed us to carry out a consistent analysis. For the first time we took Str\"{o}mgren photometry in addition to
using the 2MASS and $Gaia$ DR2 archival data. 

Our estimates for the cluster parameters such as the reddening (variable across the field, with a mean value of $E(B-V)$=0.66$\pm$0.13) and distance ($d$=2.5$\pm$0.2\,kpc)
are compatible within the errors with those found in the literature. Also the age obtained in this work ($\tau$=56$\pm$13\,Ma) is consistent with the previous values, despite being slightly
younger as a consequence of taking into account the subsolar metallicity found by us instead of assuming a solar value as it had been done so far.
We computed a tidal radius, $r_{\textrm{t}}$=18.7$\pm$1.2\,arcmin, 
significantly larger than that obtained by previous works. Nevertheless, our result is consistent with the location of the most distant members confirmed via radial velocity.
Additionally, we suspect the existence of a large halo around the cluster.

We identify 145 likely blue members from which we derive an initial mass for the cluster around 5\,200\,M$_{\sun}$. This implies that, as in the case of NGC\,3105,
NGC\,2345 is a moderately massive cluster comparable with NGC\,7419 as a template for studying obscured RSGCs and for constraining
evolutionary models of intermediate- and high-mass stars. As in NGC\,7419, we find a cluster with less than 10\,000\,M$_{\sun}$ (in this case, significantly less), 
hosting six red (super)giants, with masses around 6.5\,M$_{\sun}$. One of them, S1000 (IRAS 07055-1302), has been identified for the first time in this work.
Two other bright stars located in the halo, TYC\,5402-1851-1 and TYC\,5406-46-1, share the astrometric parameters of the cluster and, in addition, occupy a position on
the CMD compatible with that of the other red (super)giants. New spectroscopy is required to confirm their nature.
 
As NGC\,3105, NGC\,2345 is located in the outer part of the Galaxy ($R_{\textrm{GC}}$=10.2\,$\pm$\,0.2\,km\,s$^{-1}$) showing a typical LMC metallicity ([Fe/H]=$-0.28\pm0.07$).
Probably this fact is related to the high Be fraction found $\geq\,10.3\%$ among all the likely B-type members. 
 
We have studied the chemical composition of NGC\,2345 by computing the abundances of several chemical elements: Li, O, Na, $\alpha$-elements (Mg, Si, Ca and Ti);
the Fe-group (Ni) and $s$-elements (Rb, Y and Ba). We find a homogeneous composition compatible with the Galactic trends observed in the thin disc. 
The coolest cluster member, S50, is found to be a Li-rich star [A(Li)=2.11], as also happens in the case of NGC\,6067, another young open cluster.
Our results support the enhanced ``$s$-process'' model because of the overabundance of barium.

NGC\,2345, as it happens with NGC\,3105, despite its unusually low metallicity, can be considered an excellent laboratory to improve theoretical models and an optimal 
template for the study of more obscured or unresolved clusters.

Metallicity differences found along the Galactic disc affect the proper determination of the age of open clusters and, therefore, the mass of their evolved stars. 
Accordingly, metallicity must be considered as a key dimension in the study of the boundary between intermediate-mass and massive stars. Additional chemical studies of 
young open clusters will contribute  to improving our understanding of the lower mass limit for SN progenitors. 
 
\section*{Acknowledgements}

This research is based on observations collected with the MPG/ESO 2.2-meter Telescope operated at the La Silla Observatory (Chile) jointly by the Max Planck Institute 
for Astronomy and the European Organization for Astronomical Research in the Southern hemisphere under ESO programmes 087.D-603(A) and 095.A-9020(A). 
The AAT observations and the second run taken with FEROS were supported by the OPTICON project (observing proposals 2010B/01 and 2015A/008, respectively),
which is funded by the European Community's Seventh Framework Programme.
The INT and WHT telescopes are operated on the island of La Palma by the Isaac Newton Group of Telescopes in the Spanish Observatorio del Roque
de los Muchachos of the Instituto de Astrof\'{i}sica de Canarias. The WFC photometry was obtained as part of 151-INT12/11B. The IDS spectroscopy was taken as part of 32-INT3/11A.  
WYFFOS spectra were obtained under proposal 94-WHT23/12B whereas ISIS spectroscopy was made in service mode as part of SW2011b13.
This research is partially supported by the Spanish Government under grants AYA2015-68012-C2-2-P and PGC2018-093741-B-C21 (MICIU/AEI/FEDER, UE).
This research has made use of the Simbad database, operated at CDS, Strasbourg (France). This publication also made use of data products from the Two Micron All Sky Survey, which is a joint project of the 
University of Massachusetts and the Infrared Processing and Analysis Center/California Institute of Technology, funded by the National Aeronautics and Space Administration and 
the National Science Foundation. 




\bibliographystyle{aa}
\bibliography{latex} 


\appendix

\section{Additional tables}

\begin{landscape}
\begin{table}
\caption{Str\"{o}mgren photometry obtained in the field of NGC\,2345 in this work.} 
\begin{center}
\begin{tabular}{lccccccccccc}   
\hline\hline
ID  &  RA(J2000) & DEC(J2000) & $V$ & $\sigma_V$ & $(b-y)$ & $\sigma_{(b-y)}$ & $m_1$ & $\sigma_{m_1}$ & $c_1$ & $\sigma_{c_1}$  & $N$ \\
\hline
1   &  106.675384521  &  $-$13.389435768  &  16.908  &  0.020  &  0.253  &  0.027  &    1.218  &  0.045  & $-$0.800  &  0.028  & 1 \\
2   &  106.676712036  &  $-$13.385466576  &  16.514  &  0.025  &  0.556  &  0.025  &    1.055  &  0.043  & $-$0.669  &  0.030  & 1 \\
3   &  106.677474976  &  $-$13.152485847  &  16.785  &  0.017  &  0.003  &  0.025  &    1.042  &  0.039  & $-$0.254  &  0.021  & 1 \\
4   &  106.677742004  &  $-$13.337368011  &  16.028  &  0.022  &  0.508  &  0.022  &    0.976  &  0.039  & $-$0.224  &  0.025  & 1 \\
5   &  106.678443909  &  $-$13.416952133  &  15.156  &  0.050  &  0.567  &  0.059  &    0.593  &  0.065  & $-$0.030  &  0.070  & 2 \\
6   &  106.679504395  &  $-$13.219663620  &  16.263  &  0.059  &  0.779  &  0.196  & $-$0.101  &  0.224  &    0.425  &  0.242  & 2 \\
7   &  106.679794312  &  $-$13.376412392  &  14.627  &  0.036  &  0.986  &  0.032  & $-$0.074  &  0.045  &    0.519  &  0.037  & 3 \\
8   &  106.679885864  &  $-$13.230211258  &  16.126  &  0.080  &  1.071  &  0.161  &    0.247  &  0.180  &    0.037  &  0.184  & 2 \\
9   &  106.680000305  &  $-$13.412867546  &  17.148  &  0.067  &  0.443  &  0.204  &    0.622  &  0.210  & $-$0.136  &  0.210  & 3 \\
10  &  106.680206299  &  $-$13.457053185  &  16.875  &  0.044  &  0.495  &  0.105  &    0.050  &  0.103  &    1.727  &  0.100  & 2 \\

\hline
\end{tabular}
\tablefoot{Only ten rows are listed here for guidance regarding its form and content. Full table is available online at the CDS.}
\label{phot_2345}
\end{center}
\end{table}

\end{landscape}

\begin{landscape}

\begin{table}
\caption{For every star with spectroscopy are displayed the equatorial coordinates, spectral type, radial velocity (from the AA$\Omega$ spectrum), spectrographs used and cluster membership.}
\begin{center}
\begin{tabular}{lccccccccccc}   
\hline\hline
Star  &  ID & RA(J2000) & DEC(J2000) & SpT & $V_{\textrm{rad}}$ (km\,s$^{-1}$) & AA$\Omega$ & WYFFOS & FEROS & ISIS & IDS & Member\\
\hline

1    & 1239  &  107.060083333  &  $-$13.268638889 & A2\,Ib      & 66 & \textbullet  &   & \textbullet  &   & \textbullet & y\\
2    & 1148  &  107.043750000  &  $-$13.260222222 & B5\,IIIe    & 74 & \textbullet  & \textbullet  &   &   &  & y \\
4    &  970 &  107.007291667  &  $-$13.228638889 & B4\,III     & 62 & \textbullet  & \textbullet  &   & \textbullet  & \textbullet & y \\
5    & 1074  &  107.031583333  &  $-$13.222833333 & B7\,Ve      & 30 & \textbullet  &   &   &   &  & y \\
6    & 1191  &  107.051375000  &  $-$13.229638889 & B6\,V       & 56 & \textbullet  & \textbullet  &   &   &  & y \\
7    & 1265  &  107.064458333  &  $-$13.229694444 & A3\,Ib      & 56 & \textbullet  &   & \textbullet  &   & \textbullet & y \\
14   & 1320  &  107.072791667  &  $-$13.191805556 & K3\,Ib      & 59 & \textbullet  &   & \textbullet  &   &  & y \\
20   & 1195  &  107.052125000  &  $-$13.177083333 & B5\,Ve      & 61 & \textbullet  &   &   &   &  & y \\
22   & 1047  &  107.025250000  &  $-$13.186194444 & B4\,III     & 57 & \textbullet  &   &   & \textbullet  & \textbullet  & y\\
23   & 1063  &  107.029000000  &  $-$13.161388889 & B3\,V       & 57 & \textbullet  & \textbullet  &   &   &  & y \\
24   & 1171  &  107.048458333  &  $-$13.157500000 & B3\,IIIe    & 56 & \textbullet  & \textbullet  &   &   &  & y \\
25   & 1234  &  107.059708333  &  $-$13.169861111 & B3\,V       & 75 & \textbullet  &   &   &   &  & y \\
28   & 1031  &  107.069958333  &  $-$13.165750000 & B3\,IV      & 58 & \textbullet  & \textbullet  & \textbullet  & \textbullet  & \textbullet & y\\
30   & 1374  &  107.080458333  &  $-$13.116944444 & B7\,IIIe    & 54 & \textbullet  & \textbullet  &   &   &  & y \\
32   & 1385  &  107.081375000  &  $-$13.161805556 & B5\,Ve      & 55 & \textbullet  &   &   &   &  & y \\
34   & 1451  &  107.091083333  &  $-$13.173111111 & G6\,II\,+\,B& 64 & \textbullet  &   & \textbullet  &   &  & y \\
35   & 1472  &  107.095250000  &  $-$13.171888889 & B4\,IIIe    & 58 & \textbullet  & \textbullet  &   & \textbullet  & \textbullet & y \\
36   & 1767  &  107.148500000  &  $-$13.166083333 & A2\,V       &$-$6& \textbullet  & \textbullet  &   &   &   & n\\
37   & 1680  &  107.133125000  &  $-$13.169944444 & B3\,IV      & 61 & \textbullet  &   &   &   &  & y \\
39   & 1723  &  107.140375000  &  $-$13.182750000 & B5\,IV      & 56 & \textbullet  & \textbullet  &   &   &  & y \\
43   & 1553  &  107.109687500  &  $-$13.187344444 & K0\,Ib-II   & 58 & &   & \textbullet  &   &  & y \\
44   & 1526  &  107.106291667  &  $-$13.200666667 & B5\,Ve      & 52 & \textbullet  & \textbullet  &   &   &  & y \\
47   & 1623  &  107.122000000  &  $-$13.212591667 & B4\,V       & 56 & &   &   & \textbullet  & \textbullet  & n\\
50   & 1568  &  107.112541667  &  $-$13.209166667 & K4\,Ib      & 59 & \textbullet  &   & \textbullet  &   &  & y \\
51   & 1558  &  107.109875000  &  $-$13.210166667 & B3\,IV      & 58 & \textbullet  & \textbullet  &   & \textbullet  & \textbullet & y \\
54   & 1506  &  107.103041667  &  $-$13.229388889 & B5\,V       & 43 & \textbullet  & \textbullet  &   &   &   & y\\
56   & 1544  &  107.108666667  &  $-$13.240111111 & B6\,V       & 49 & \textbullet  &   &   &   &   & y\\
59   & 1595  &  107.116791667  &  $-$13.259805556 & B5\,Ve      & 54 & \textbullet  &   &   &   &   & y\\
60   & 1646  &  107.126541667  &  $-$13.231250000 & K2\,Ib-II   & 58 & \textbullet  &   & \textbullet  &   &  & y \\
61   & 1640  &  107.124791667  &  $-$13.220888889 & B7\,Ve      & 50 & \textbullet  & \textbullet  &   &   &   & y\\
63   &       &  107.294583333  &  $-$13.181111111 & B5\,Ve      & 44 & \textbullet  & \textbullet  &   &   &   & y\\
64   &  899  &  106.988375000  &  $-$13.058833333 & B3\,IV      & 59 & \textbullet  & \textbullet  &   &   &  & y \\
1000 &  873  &  106.982500000  &  $-$13.128416667 & K4-5\,Ib    & 59 & \textbullet  &   & \textbullet  &   &  & y\\
1002 &       &  107.297083333  &  $-$13.434694444 & M6\,II      &    & \textbullet  &   & \textbullet  &   &  & n\\
1003 &  811  &  106.964625000  &  $-$13.076305556 & B5\,Ve      & 48 & \textbullet  & \textbullet  &   &   &  & y\\
1004 &       &  107.575791667  &  $-$12.703333333 & B3\,V       & 32 & \textbullet  &   &   &   &   & n\\
1005 &       &  107.487541667  &  $-$13.616055556 & G4\,II      &    & \textbullet  &   &   &   &  & n \\
1009 &       &  106.442000000  &  $-$13.160305556 & B0.5e       &    & \textbullet  & \textbullet  &   &   &   & n\\
1010 &       &  107.355875000  &  $-$12.541916667 & BN2\,Ib     & 49 & \textbullet  &   &   &   &   & n\\
1011 & 1387  &  107.081083333  &  $-$12.980361111 & F\,V:       & 40 & \textbullet  & \textbullet  &   &   &  & n \\
1012 &  902  &  106.990291667  &  $-$13.161388889 & B3\,V       & 47 & \textbullet  & \textbullet  &   &   &  & y\\
1020 &  586  &  106.901708333  &  $-$13.264388889 & B8\,III     & 59 & \textbullet  & \textbullet  &   &   &   & n\\

\hline
\end{tabular}

\label{spectra_2345}
\end{center}
\end{table}

\end{landscape}

\begin{landscape}

\begin{table}
\contcaption{}
\begin{center}
\begin{tabular}{lcccccccccccc}   
\hline\hline
Star  &  ID & RA(J2000) & DEC(J2000) & SpT & $V_{\textrm{rad}}$ (km\,s$^{-1}$) & AA$\Omega$ & WYFFOS & FEROS & ISIS & IDS & Member\\
\hline
1021     &  427 &  106.855083333  &  $-$13.295472222 & late A\,III & 39 & \textbullet  & \textbullet  &   &   &   & n\\
1022     &  981 &  107.010583333  &  $-$13.436277778 & B3\,V       & 31 & \textbullet  & \textbullet  &   &   &  & n \\
1023     &      &  107.355958333  &  $-$12.926277778 & B5\,V       & 12 & \textbullet  & \textbullet  &   &   &  & n \\
1024     &      &  107.288833333  &  $-$12.962666667 & late A\,III & 36 & \textbullet  & \textbullet  &   &   &  & n \\
1025     &      &  107.261416667  &  $-$12.961500000 & A3\,V       & 14 & \textbullet  & \textbullet  &   &   &  & n \\
1026     &      &  107.308041667  &  $-$12.942138889 & early A     & 21 & \textbullet  &   &   &   &   & n\\
1027     &      &  107.327375000  &  $-$13.042611111 & A5\,V       & 66 & \textbullet  & \textbullet  &   &   &  & n\\
1028     &      &  107.385375000  &  $-$13.076972222 & A0\,III     & 22 & \textbullet  & \textbullet  &   &   &   & n\\
1029     &      &  107.375083333  &  $-$13.127833333 & A7\,II-III  & 23 & \textbullet  & \textbullet  &   &   &   & n\\
1030     &      &  106.788125000  &  $-$12.978861111 & A2\,III     & 15 & \textbullet  & \textbullet  &   &   &  & n \\
1031     &      &  106.809375000  &  $-$12.979916667 & B9.5\,II-III& 65 & \textbullet  & \textbullet  &   &   &  & n\\
1032     &  306 &  106.803875000  &  $-$13.216527778 & A1\,V       & 30 & \textbullet  & \textbullet  &   &   &  & n \\
1034     & 1704 &  107.136500000  &  $-$12.986277778 & early F     &  0 & \textbullet  & \textbullet  &   &   &  & n \\
1035     & 1539 &  107.107791667  &  $-$13.004750000 & A0\,V       & 13 & \textbullet  & \textbullet  &   &   &  & n \\
1050     &      &  107.602041667  &  $-$13.564472222 & A:          & 36 & \textbullet  &   &   &   &  & n \\
1051     &      &  107.149833333  &  $-$13.989472222 & B5\,III\,shell& 78 & \textbullet  &   &   &   &  & n \\
1125     &      &  106.382750000  &  $-$13.401666667 & B5\,III     & 37 & \textbullet  &   &   &   &  & n \\
1126     &      &  106.368125000  &  $-$13.353416667 & B8\,III     & 15 & \textbullet  &   &   &   &  & n \\
LS\,152  &      &  106.406416667  &  $-$13.523333333 & B2\,IV    & 64 & \textbullet  &   &   &   &   & n\\
LS\,153  &      &  106.404791667  &  $-$13.775888889 & B0.7\,III & 91 & \textbullet  &   &   &   &  & n \\
LS\,164  &      &  106.556541667  &  $-$13.185861111 & O8\,V((f))z& 70& \textbullet  & \textbullet  &   &   &   & n\\
LS\,165  &      &  106.605083333  &  $-$13.800444444 & B0.7\,III &    & \textbullet  &   &   &   &  & n \\
LS\,170  &   86 &  106.715083333  &  $-$13.287694444 & B9\,Ib    & 99 & \textbullet  & \textbullet  &   &   &   & n\\
LS\,171  &      &  106.734958333  &  $-$12.774555556 & B1.5\,Ve  & 71 & \textbullet  & \textbullet  &   &   &  & n\\
LS\,174  &      &  106.813041667  &  $-$13.659750000 & B2\,Ve    & 57 & \textbullet  & \textbullet  &   &   &   & y?\\
LS\,176  &      &  106.955375000  &  $-$12.787944444 & B3\,IIIe  & 65 & \textbullet  & \textbullet  &   &   &  & y?\\
LS\,180  &      &  107.029333333  &  $-$12.805333333 & B5\,V     & 62 & \textbullet  & \textbullet  &   &   &  & n\\
LS\,193  & 2093 &  107.242875000  &  $-$13.432277778 & B3\,IV & 47 & \textbullet  & \textbullet  & \textbullet  &   &   & y\\
LS\,205  &      &  107.489291667  &  $-$13.502833333 & B3\,III   & 72 & \textbullet  &   &   &   &   & y?\\
LS\,215  &      &  107.646958333  &  $-$13.053305556 &           & 42 & \textbullet  &   &   &   &   & n\\
LS\,221  &      &  107.748666667  &  $-$13.526222222 &           &    & \textbullet  &   &   &   &   & n\\
LS\,224  &      &  107.809166667  &  $-$13.510833333 &           & 33 & \textbullet  &   &   &   &   & n\\
LS\,229  &      &  107.885583333  &  $-$12.844277778 & B2.5\,III & 72 & \textbullet  &   &   &   &   & y?\\
HD\,54743  &    &  107.373333333  &  $-$13.176000000 & G5\,III   &  8 & \textbullet  &   &   &   &   & n\\

\hline
\end{tabular}

\label{spectra_2345}
\end{center}
\end{table}

\end{landscape}

\begin{table*}
\caption{$Gaia$ DR2 astrometric data for all stars in the field of NGC\,2345 for which we have spectra and other objects discussed
in the text.}
\label{tab_mp}
\begin{center}
\begin{tabular}{lccc}   
\hline\hline
Star    & $\varpi$ (mas)  &  $\mu_{\alpha*}$ (mas\,a$^{-1}$) & $\mu_{\delta}$ (mas\,a$^{-1}$) \\
\hline
\multicolumn{4}{c}{Members}\\    
\hline
1    &   0.3615 $\pm$ 0.0342  &  $-$1.537 $\pm$ 0.054  & 1.472 $\pm$ 0.048  \\
2    &   0.3507 $\pm$ 0.0330  &  $-$1.413 $\pm$ 0.050  & 1.351 $\pm$ 0.045  \\  
4    &   0.3904 $\pm$ 0.0343  &  $-$1.360 $\pm$ 0.052  & 1.415 $\pm$ 0.046  \\
5    &   0.3464 $\pm$ 0.0214  &  $-$1.392 $\pm$ 0.031  & 1.409 $\pm$ 0.028  \\  
6    &   0.3154 $\pm$ 0.0214  &  $-$1.213 $\pm$ 0.032  & 1.313 $\pm$ 0.029  \\
7    &   0.3906 $\pm$ 0.0361  &  $-$1.446 $\pm$ 0.052  & 1.248 $\pm$ 0.046  \\
14   &   0.4448 $\pm$ 0.0592  &  $-$1.692 $\pm$ 0.087  & 1.536 $\pm$ 0.079  \\
20   &   0.3261 $\pm$ 0.0232  &  $-$1.321 $\pm$ 0.032  & 1.287 $\pm$ 0.030  \\
22   &   0.3687 $\pm$ 0.0339  &  $-$1.355 $\pm$ 0.052  & 1.330 $\pm$ 0.047  \\
23   &   0.3615 $\pm$ 0.0216  &  $-$1.401 $\pm$ 0.030  & 1.325 $\pm$ 0.027  \\
24   &   0.3671 $\pm$ 0.0226  &  $-$1.390 $\pm$ 0.032  & 1.310 $\pm$ 0.029  \\
25   &   0.3723 $\pm$ 0.0234  &  $-$1.242 $\pm$ 0.032  & 1.236 $\pm$ 0.029  \\
27   &   0.3170 $\pm$ 0.0239  &  $-$1.378 $\pm$ 0.032  & 1.322 $\pm$ 0.029  \\
28   &   0.3653 $\pm$ 0.0363  &  $-$1.336 $\pm$ 0.050  & 1.270 $\pm$ 0.045  \\
30   &   0.3675 $\pm$ 0.0246  &  $-$1.341 $\pm$ 0.034  & 1.309 $\pm$ 0.032  \\
32   &   0.3378 $\pm$ 0.0234  &  $-$1.353 $\pm$ 0.032  & 1.324 $\pm$ 0.029  \\
34   &   0.3670 $\pm$ 0.0390  &  $-$1.201 $\pm$ 0.055  & 1.238 $\pm$ 0.050  \\
35   &   0.2823 $\pm$ 0.0372  &  $-$1.464 $\pm$ 0.052  & 1.387 $\pm$ 0.048  \\
37   &   0.3930 $\pm$ 0.0330  &  $-$1.284 $\pm$ 0.048  & 1.357 $\pm$ 0.044  \\
39   &   0.3335 $\pm$ 0.0322  &  $-$1.327 $\pm$ 0.048  & 1.342 $\pm$ 0.043  \\
43   &   0.3926 $\pm$ 0.0351  &  $-$1.241 $\pm$ 0.050  & 1.238 $\pm$ 0.045  \\
44   &   0.3465 $\pm$ 0.0241  &  $-$1.280 $\pm$ 0.032  & 1.295 $\pm$ 0.028  \\
50   &   0.3386 $\pm$ 0.0434  &  $-$1.273 $\pm$ 0.063  & 1.179 $\pm$ 0.058  \\
51   &   0.3321 $\pm$ 0.0350  &  $-$1.413 $\pm$ 0.051  & 1.505 $\pm$ 0.053  \\
54   &   0.3239 $\pm$ 0.0404  &  $-$1.293 $\pm$ 0.058  & 1.265 $\pm$ 0.049  \\
56   &   0.3234 $\pm$ 0.0241  &  $-$1.425 $\pm$ 0.035  & 1.230 $\pm$ 0.032  \\
59   &   0.3431 $\pm$ 0.0281  &  $-$1.360 $\pm$ 0.042  & 1.489 $\pm$ 0.037  \\
60   &   0.3457 $\pm$ 0.0380  &  $-$1.403 $\pm$ 0.051  & 1.431 $\pm$ 0.046  \\
61   &   0.3407 $\pm$ 0.0320  &  $-$1.336 $\pm$ 0.042  & 1.384 $\pm$ 0.037  \\
63   &   0.3558 $\pm$ 0.0335  &  $-$1.295 $\pm$ 0.051  & 1.336 $\pm$ 0.046  \\
64   &   0.3564 $\pm$ 0.0427  &  $-$1.382 $\pm$ 0.060  & 1.345 $\pm$ 0.052  \\
1000 &   0.3932 $\pm$ 0.0539  &  $-$1.712 $\pm$ 0.078  & 0.950 $\pm$ 0.066  \\
1003 &   0.3517 $\pm$ 0.0376  &  $-$1.437 $\pm$ 0.056  & 1.328 $\pm$ 0.050  \\
1012 &   0.3796 $\pm$ 0.0273  &  $-$1.337 $\pm$ 0.049  & 1.269 $\pm$ 0.039  \\
LS\,193& 0.3786 $\pm$ 0.0386  &  $-$1.231 $\pm$ 0.059  & 1.356 $\pm$ 0.050  \\
X1   &   0.3481 $\pm$ 0.0221  &  $-$1.343 $\pm$ 0.033  & 1.359 $\pm$ 0.030  \\
X2   &   0.3531 $\pm$ 0.0236  &  $-$1.276 $\pm$ 0.032  & 1.287 $\pm$ 0.028  \\
\hline
\multicolumn{4}{c}{Halo members}\\    
\hline
LS\,174          & 0.3070 $\pm$ 0.0415  &  $-$1.253 $\pm$ 0.055  & 1.578 $\pm$ 0.056  \\
LS\,176          & 0.2792 $\pm$ 0.0330  &  $-$1.228 $\pm$ 0.062  & 1.278 $\pm$ 0.058  \\
LS\,205          & 0.2749 $\pm$ 0.0328  &  $-$1.129 $\pm$ 0.045  & 1.220 $\pm$ 0.041  \\
LS\,229          & 0.2906 $\pm$ 0.0564  &  $-$1.182 $\pm$ 0.083  & 1.037 $\pm$ 0.077  \\
TYC-5402-1851-1  & 0.3168 $\pm$ 0.0479  &  $-$1.067 $\pm$ 0.066  & 1.290 $\pm$ 0.059  \\
TYC-5406-46-1    & 0.2941 $\pm$ 0.0406  &  $-$1.424 $\pm$ 0.084  & 1.254 $\pm$ 0.076  \\
\hline
\multicolumn{4}{c}{Non members}\\    
\hline
36   &   1.0756 $\pm$ 0.0350  &  $+$0.525 $\pm$ 0.051  & 3.257 $\pm$ 0.046  \\
47   &   0.0802 $\pm$ 0.1666  &  $+$2.770 $\pm$ 0.259  & $-$3.205 $\pm$ 0.222  \\
1002 &   0.4178 $\pm$ 0.1514  &  $-$2.556 $\pm$ 0.228  & 1.936 $\pm$ 0.196  \\
1004 &   0.6156 $\pm$ 0.0601  &  $-$1.781 $\pm$ 0.094  & $-$1.044 $\pm$ 0.090  \\
1005 &   2.1214 $\pm$ 0.0507  &  $-$3.566 $\pm$ 0.070  & $-$7.884 $\pm$ 0.061  \\
1009 &   0.0537 $\pm$ 0.0383  &  $-$0.152 $\pm$ 0.064  &    0.780 $\pm$ 0.064  \\
1010 &   0.3709 $\pm$ 0.0475  &  $-$1.240 $\pm$ 0.070  &    1.481 $\pm$ 0.073  \\
1011 &   1.0805 $\pm$ 0.0372  &  $-$1.817 $\pm$ 0.057  &    4.019 $\pm$ 0.052  \\
1020 &   0.3252 $\pm$ 0.0257  &  $-$0.620 $\pm$ 0.038  &    0.364 $\pm$ 0.036  \\
1021 &   0.9576 $\pm$ 0.0426  &  $-$4.399 $\pm$ 0.074  &    0.117 $\pm$ 0.069  \\
1022 &   0.5039 $\pm$ 0.0382  &  $-$2.033 $\pm$ 0.062  &    1.170 $\pm$ 0.056  \\
1023 &   0.4986 $\pm$ 0.0415  &  $-$1.908 $\pm$ 0.061  &    1.130 $\pm$ 0.055  \\
1024 &   0.8040 $\pm$ 0.0329  &  $-$3.270 $\pm$ 0.053  &    2.525 $\pm$ 0.046  \\

\hline
\end{tabular}
\end{center}
\end{table*}

\begin{table*}
\contcaption{}
\begin{center}
\begin{tabular}{lccc}   
\hline\hline
Star     & $\varpi$ (mas)  &  $\mu_{\alpha*}$ (mas\,a$^{-1}$) & $\mu_{\delta}$ (mas\,a$^{-1}$) \\
\hline
1025     &   0.7865 $\pm$ 0.0227  &  $-$3.007 $\pm$ 0.035  &    0.544 $\pm$ 0.033  \\
1026     &   0.3746 $\pm$ 0.0224  &     0.019 $\pm$ 0.036  & $-$0.508 $\pm$ 0.031  \\
1027     &   1.1490 $\pm$ 0.0329  &  $-$3.297 $\pm$ 0.052  &    3.115 $\pm$ 0.046  \\
1028     &   0.9284 $\pm$ 0.0335  &  $-$3.725 $\pm$ 0.051  &    0.382 $\pm$ 0.045  \\
1029     &   1.2657 $\pm$ 0.0411  &  $-$4.539 $\pm$ 0.072  &    2.401 $\pm$ 0.067  \\
1030     &$-$1.9468 $\pm$ 0.5428  &     1.207 $\pm$ 1.042  & $-$4.808 $\pm$ 0.986  \\
1031     &   0.4076 $\pm$ 0.0222  &  $-$0.548 $\pm$ 0.035  & $-$0.121 $\pm$ 0.037  \\
1032     &   0.3987 $\pm$ 0.0395  &  $-$1.670 $\pm$ 0.054  &    0.007 $\pm$ 0.052  \\
1034     &   1.1756 $\pm$ 0.0398  &     0.869 $\pm$ 0.061  &    1.108 $\pm$ 0.054  \\
1035     &   0.9189 $\pm$ 0.0385  &  $-$3.373 $\pm$ 0.060  &    0.694 $\pm$ 0.052  \\
1050     &   0.7041 $\pm$ 0.0402  &  $-$0.970 $\pm$ 0.061  & $-$0.118 $\pm$ 0.054  \\
1051     &   4.1798 $\pm$ 0.0641  &  $-$2.717 $\pm$ 0.086  & $-$4.901 $\pm$ 0.076  \\
1125     &   0.3853 $\pm$ 0.0511  &  $-$2.040 $\pm$ 0.077  &    1.022 $\pm$ 0.074  \\
1126     &   0.6767 $\pm$ 0.0507  &  $-$0.663 $\pm$ 0.069  &    0.131 $\pm$ 0.067  \\
LS\,152  &   0.3871 $\pm$ 0.0394  &  $-$1.068 $\pm$ 0.067  &    0.269 $\pm$ 0.065  \\
LS\,153  &   0.0337 $\pm$ 0.0354  &  $-$0.321 $\pm$ 0.058  &    0.804 $\pm$ 0.053  \\
LS\,164  &$-$0.0236 $\pm$ 0.0612  &  $-$0.396 $\pm$ 0.076  &    0.852 $\pm$ 0.085  \\
LS\,165  &   0.0732 $\pm$ 0.0415  &  $-$0.421 $\pm$ 0.064  &    0.958 $\pm$ 0.059  \\
LS\,170  &   0.0948 $\pm$ 0.1013  &     0.029 $\pm$ 0.139  &    0.881 $\pm$ 0.153  \\
LS\,171  &   0.0590 $\pm$ 0.0349  &  $-$0.250 $\pm$ 0.062  &    1.553 $\pm$ 0.057  \\
LS\,180  &   0.3845 $\pm$ 0.0447  &  $-$2.089 $\pm$ 0.069  &    2.350 $\pm$ 0.065  \\
LS\,215  &   0.4011 $\pm$ 0.0385  &  $-$0.861 $\pm$ 0.056  &    1.268 $\pm$ 0.059  \\
LS\,221  &   0.1266 $\pm$ 0.0335  &  $-$1.053 $\pm$ 0.050  &    0.998 $\pm$ 0.046  \\
LS\,224  &   0.2552 $\pm$ 0.0289  &  $-$0.902 $\pm$ 0.044  &    1.112 $\pm$ 0.040  \\
HD\,54743  & 1.5844 $\pm$ 0.0381  &  $-$9.664 $\pm$ 0.056  & $-$0.122 $\pm$ 0.050  \\
TV\,CMa  &   0.3137 $\pm$ 0.0335  &  $-$0.743 $\pm$ 0.050  & 0.104 $\pm$ 0.045  \\
ASAS\,J070911-1217.2  &   0.3140 $\pm$ 0.0564  &  $-$1.045 $\pm$ 0.082  & 1.276 $\pm$ 0.091  \\

\hline
\end{tabular}
\end{center}
\end{table*}

\begin{table*}
\caption{Photometric and astrometric data for likely blue members of NGC\,2345.}
\begin{center}
\begin{tabular}{lcccccccccc} 
\hline\hline
ID & $E(B-V)$ & $V$ & $(b-y)$ & $J$ & $(J-K_{\textrm{S}})$ & $G$ & $(G_{\textrm{BP}}-G_{\textrm{RP}})$ & $\varpi$ (mas)  &  $\mu_{\alpha*}$ (mas\,a$^{-1}$) & $\mu_{\delta}$ (mas\,a$^{-1}$) \\
\hline
131  & 0.757  & 15.973  & 0.549  & 14.442  & 0.434  & 15.789  & 0.980  & 0.422  & $-$1.419  & 1.424 \\ 
137  & 0.576  & 14.363  & 0.380  & 13.167  & 0.296  & 14.222  & 0.705  & 0.345  & $-$1.336  & 1.439 \\ 
140  & 0.849  & 16.038  & 0.604  & 14.317  & 0.461  & 15.762  & 1.078  & 0.389  & $-$1.113  & 1.235 \\ 
169  & 1.130  & 15.273  & 0.784  & 13.017  & 0.435  & 14.916  & 1.426  & 0.318  & $-$1.188  & 1.408 \\ 
242  & 0.480  & 14.311  & 0.326  & 13.346  & 0.139  & 14.234  & 0.591  & 0.290  & $-$1.411  & 1.436 \\ 
243  & 0.859  & 16.133  & 0.613  & 14.345  & 0.535  & 15.939  & 1.172  & 0.420  & $-$1.387  & 1.310 \\ 
248  & 0.769  & 14.916  & 0.547  & 13.494  & 0.344  & 14.785  & 0.957  & 0.371  & $-$1.697  & 1.203 \\ 
307  & 0.757  & 15.624  & 0.542  & 14.383  & 0.334  & 15.566  & 0.824  & 0.250  & $-$1.362  & 1.563 \\ 
317  & 0.986  & 16.454  & 0.717  & 14.512  & 0.446  & 16.162  & 1.242  & 0.316  & $-$1.332  & 1.121 \\ 
353  & 0.599  & 14.515  & 0.394  & 13.248  & 0.258  & 14.361  & 0.801  & 0.352  & $-$1.440  & 1.336 \\ 
\hline
\end{tabular}
\tablefoot{Only ten rows are listed here for guidance regarding its form and content. Full table is available online at the CDS.}
\label{phot_spt}
\end{center}
\end{table*}


\end{document}